\documentclass[prd,a4paper,twocolumn,fleqn,groupedaddress,nofootinbib,preprintnumbers]{revtex4-1}

\usepackage{graphicx,color}
\usepackage{amsmath,amssymb,amsfonts}

\newcommand{\be}{\begin{equation}}
\newcommand{\ee}{\end{equation}}

\newcommand{\bea}{\begin{eqnarray}}
\newcommand{\eea}{\end{eqnarray}}

\newcommand{\ba}{\begin{eqnarray}}
\newcommand{\ea}{\end{eqnarray}}

\let\Im\relax
\DeclareMathOperator{\Im}{Im}

\def\be{\begin{equation}}
\def\ee{\end{equation}}
\def\beq{\begin{eqnarray}}
\def\eeq{\end{eqnarray}}

\def\d{\delta}

\begin{document}

\input amssym.def
\input amssym.tex

\title{Thermalization of Wightman functions in AdS/CFT and quasinormal modes}

\preprint{OUTP-15-29P}

\author{Ville Ker\"{a}nen}
\email[E-mail: ]{vkeranen1@gmail.com}
\author{Philipp Kleinert}
\email[E-mail: ]{philipp.kleinert@physics.ox.ac.uk}
\affiliation{Rudolf Peierls Centre for Theoretical Physics, University of Oxford,\\ 1 Keble Road, Oxford OX1 3NP, United Kingdom}

\begin{abstract}
We study the time evolution of Wightman two-point functions of scalar fields in AdS$_3$-Vaidya, a spacetime undergoing gravitational collapse. In the boundary field theory, the collapse corresponds to a quench process where the dual 1+1 dimensional CFT is taken out of equilibrium and subsequently thermalizes. From the two-point function, we extract an effective occupation number in the boundary theory and study how it approaches the thermal Bose-Einstein distribution. We find that the Wightman functions, as well as the effective occupation numbers, thermalize with a rate set by the lowest quasinormal mode of the scalar field in the BTZ black hole background. We give a heuristic argument for the quasinormal decay, which is expected to apply to more general Vaidya spacetimes also in higher dimensions. This suggests a unified picture in which thermalization times of one- and two-point functions are determined by the lowest quasinormal mode. Finally, we study how these results compare to previous calculations of two-point functions based on the geodesic approximation.
\end{abstract}

\maketitle

\section{Introduction}

The AdS/CFT duality provides a novel tool for the study of nonequilibrium real time dynamics of strongly coupled quantum field theories in terms of semiclassical gravity. An interesting question to study in nonequilibrium quantum field theory is how fast (if at all) a field theory relaxes towards thermal equilibrium after it is perturbed out of equilibrium. This process is called thermalization. The time scale at which a given quantity approaches the value that it takes in a thermal state (with temperature determined by the energy of the initial nonequilibrium state) is called the thermalization time associated with the observable. In many situations, in the context of holography, it has been found that expectation values of local operators, which we will call one-point functions, approach their thermal values with rates dictated by  the lowest quasinormal mode of the corresponding bulk field (see \cite{Bhaseen:2012gg,Craps:2015upq} for two examples demonstrating this point). Phrased in field theory language, the time scale on which one-point functions thermalize, starting from a far-from-equilibrium state, is the same on which infinitesimally small perturbations away from thermal equilibrium decay back to equilibrium.

On the other hand, the thermalization time scale of nonlocal observables, such as two-point functions of local operators, spacelike Wilson loops and entanglement entropy\ \cite{AbajoArrastia:2010yt,Albash:2010mv,Balasubramanian:2011ur,Liu:2013qca,Buchel:2014gta,Ecker:2015kna} depends crucially on the length scale $l$ associated with the observable, such as the distance between the points of the two point correlation function. In many cases, the thermalization time increases linearly with the length scale, $t_{therm}\propto l$.  Thus, it seems that the thermalization time scale of the nonlocal observables behaves very differently from that of local observables. Our main claim in this paper is that there is actually no difference between the thermalization time scales of local and nonlocal observables, at least for two point correlation functions in momentum space. We will provide evidence that the two-point functions, when being Fourier transformed with respect to their spatial coordinates, approach the thermal two-point functions with a rate set by the lowest quasinormal mode in thermal equilibrium. This result suggests a unified picture of thermalization time scales being determined by quasinormal modes.

Our analysis mainly focuses on AdS$_3$-Vaidya spacetime,
\beq
	ds^2\!=\!\frac{L^2}{z^2}\Big[\!-\![1\!-\!\theta(v)(2\pi T_f)^2z^2]dv^2\!-\!2dv dz\!+\!dx^2\Big]\!,\,\,\,\,
	\label{eq:metric}
\eeq
which provides a simple example of a collapsing spacetime. It has been argued to be a reasonably good approximation to many more realistic out of equilibrium states \cite{Bhattacharyya:2009uu,Wu:2012rib,Garfinkle:2011tc,Horowitz:2013mia}. The coordinate $v$ in (\ref{eq:metric}) is a null time coordinate that reduces at the boundary to the boundary theory time coordinate denoted by $t$. In the dual field theory, the spacetime (\ref{eq:metric}) corresponds to initializing the field theory in the vacuum state at times $t\!<\!0$. At $t\!=\!0$, the field theory is kicked out of equilibrium which is manifested on the AdS side by a shell of matter that falls along a lightlike geodesic and collapses into a black hole. The Hawking temperature of the final black hole state is denoted as $T_f$ and can be identified as the field theory temperature once thermal equilibrium has been reached at late times. In Vaidya spacetime, the expectation value of the energy momentum tensor becomes immediately thermal after $t\!=\!0$, while the two point correlation functions take more time to thermalize. Thus, even though the black hole forms suddenly at $v\!=\!0$ in the null coordinate system, nonlocal observables in the field theory have memory of the nonthermal initial state.

A question worth addressing is why one should consider two point correlation functions at all? As a motivation, we recall some of the applications of Wightman two-point functions. Firstly, while one-point functions in a quantum system give average values of observable quantities, Wightman two-point functions quantify fluctuations of the value of the observable in question.\footnote{E.g.\ the measured values of an observable $A$ has the variance $\Delta A^2=\langle A^2\rangle-\langle A\rangle^2$.} Secondly, Wightman two-point functions quantify particle production and absorption rates. For example, the photon production rate of the quark-gluon plasma is proportional to the current operator Wightman two-point function at leading order in the electromagnetic coupling (see e.g.\ \cite{Bellac:2011kqa}). Finally, an occupation number for out of equilibrium physics in the context of holography was introduced in \cite{Balasubramanian:2012tu}. The computation of this quantity requires the knowledge of the Wightman two-point function.\footnote{Originally in \cite{Balasubramanian:2012tu}, the occupation number was written in terms of the Feynman and the retarded correlator, but the same formula can be written in a simpler way in terms of the Wightman function.} The definition was inspired by the fluctuation dissipation relation in a thermal state and has already been used earlier in the context of nonequilibrium quantum field theory (see e.g.\ \cite{Berges:2010ez}). It is well defined even out of equilibrium and without well-defined particle states. In the following, we will also calculate this quantity in AdS$_3$-Vaidya and study how it approaches the thermal Bose-Einstein distribution.

This paper is organized as follows. In Section \ref{sec:1}, we briefly review how scalar two-point functions can be calculated by solving the bulk Klein-Gordon equation and using the ``extrapolate" dictionary. In Section \ref{sec:2}, we present numerical results for the boundary theory Wightman function in AdS$_3$-Vaidya spacetime for a scalar field corresponding to an operator with scaling dimension $\Delta=3/2$ and show numerical evidence for the quasinormal decay of the correlation functions towards their thermal values. In Section \ref{sec:3}, we provide an explanation for the quasinormal mode decay that is expected to apply to AdS-Vaidya spacetimes in higher dimensions. We substantiate this explanation by a concrete computation in the case of  AdS$_3$-Vaidya. In Section \ref{sec:4}, we present a simpler numerical method for calculating nonequilibrium two-point functions in AdS-Vaidya spacetime inspired by the heuristic argument of the previous section. In Section \ref{sec:5}, we extract an effective occupation number from the boundary Wightman function and study how it approaches the thermal Bose-Einstein distribution. In Section \ref{sec:6}, we present numerical evidence that the geodesic approximation also leads to quasinormal decay towards thermal equilibrium in AdS$_3$-Vaidya spacetime. In Section \ref{sec:7}, we present our conclusions.

\section{Numerical calculation of the Vaidya Wightman functions}\label{sec:1}

We will study two point correlation functions of a minimally coupled scalar field $\phi$ with mass $m^2=-3/4$. In the ``extrapolate" version of the AdS/CFT dictionary \cite{Banks:1998dd,Giddings:1999qu}, the boundary theory Wightman function is given by
\begin{align}
G_{\pm}^{CFT}&(x_1,x_2)\notag\\
	=\,& 2\pi\!\!\lim_{z_1,z_2\rightarrow 0}(z_1z_2)^{-\Delta}G_{\pm}(x_1,z_1;x_2,z_2),\label{eq:boundary_correlator}
\end{align}
where the correlator on the right hand side is one of the bulk Wightman functions
\begin{align}
G_+(x_1,z_1;x_2,z_2)=\langle \phi(x_1,z_1)\phi(x_2,z_2)\rangle\nonumber,
\\
G_-(x_1,z_1;x_2,z_2)=\langle \phi(x_2,z_2)\phi(x_1,z_1)\rangle.
\end{align}
Another version of the out-of-equilibrium dictionary in AdS/CFT is the Skenderis-van Rees dictionary \cite{Skenderis:2008dg}, where one constructs a bulk spacetime for the Schwinger-Keldysh contour. In\ \cite{Keranen:2014lna}, we showed that these two dictionaries are identical at the level of scalar two-point functions. Thus, it does not matter which one we use here.

The bulk Wightman functions satisfy the Klein-Gordon equation with respect to both of their arguments,
\begin{align}
&(\Box_1-m^2)G_{\pm}(x_1,z_1;x_2,z_2)=0,\nonumber
\\
&(\Box_2-m^2)G_{\pm}(x_1,z_1;x_2,z_2)=0.\label{eq:eoms}
\end{align}
The initial data for solving these equations is fixed by the correlations in the initial state of the scalar field. In AdS-Vaidya spacetime, a natural initial state for the bulk scalar is the AdS ground state. This corresponds to the dual field theory being prepared in the vacuum state. The main difficulty in solving (\ref{eq:eoms}) is that the correlators have singularities at short Lorentzian distances. A numerical strategy for solving the differential equations, based on the method of Green's functions, was presented in detail in our previous paper \cite{Keranen:2014lna}.  The basic method of calculation in \cite{Keranen:2014lna} is reviewed in Appendix \ref{sec:method_old}. In Section \ref{sec:4}, we will sketch a simpler way of solving the problem. The details of this method are laid out in Appendix~\ref{sec:method_new}. We have compared the two methods and find that they give the same results within the limits of numerical accuracy. Some of the plots in this paper are made using this new method, which allows us to evolve to much later times compared to the previous paper.

\section{Results for the boundary theory Wightman functions}\label{sec:2}

\begin{figure}[t]
\begin{center}
\includegraphics[width=0.975\linewidth]{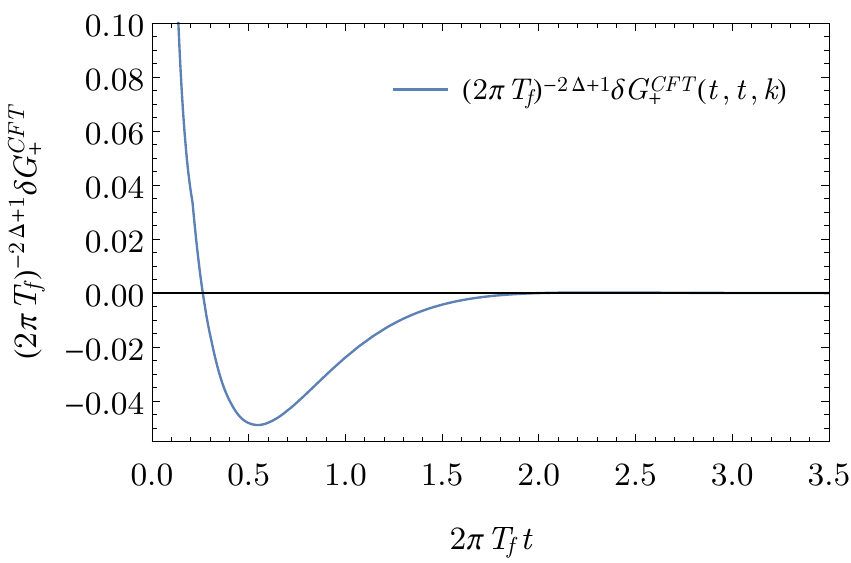}\\ \vspace{0.5cm}
\includegraphics[width=0.975\linewidth]{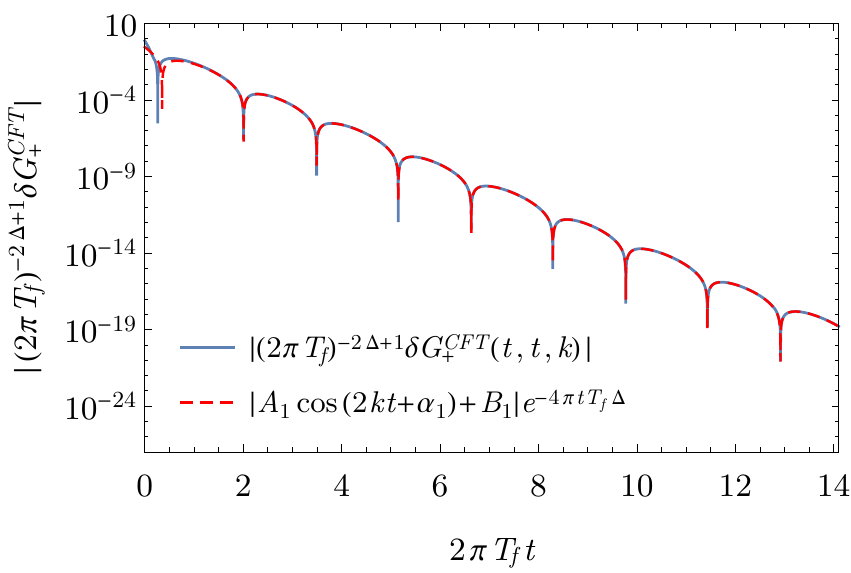}
\caption{\label{fig:Decayfig} 
The boundary theory equal-time Wightman function as a function of time with the thermal Wightman function subtracted, as defined in (\ref{eq:subtracted}). The blue solid curves correspond to the value of the correlator while the red dashed curve is a fit to the lowest quasinormal mode of the BTZ black hole in the form (\ref{eq:quasi}). We have fixed $k=2\pi T_f$.}
\end{center}
\end{figure}

In what follows, we will consider the boundary theory two-point function after performing a Fourier transform with respect to the boundary spatial coordinate $x$
\beq
	G^{CFT}_+(t_1,t_2;k)=\int_{-\infty}^{\infty}\!\!\!\! dx\,e^{i kx}\,G_+^{CFT}(t_1,x;t_2,0).\quad
\eeq
The solid blue curve in Fig.\ \ref{fig:Decayfig} shows how the subtracted boundary Wightman function
\begin{equation}
	\delta G_+^{CFT}=G_+^{CFT}-G_+^{CFT(thermal)},\label{eq:subtracted}
\end{equation}
evolves in time as both of its time arguments $t_1$ and $t_2$ are set equal and then evolved forwards. The spatial momentum is chosen to take the value $k\!=\!2\pi T_f$. Similar results have been obtained for other values of $k$. The Wightman function approaches thermal equilibrium in a damped oscillating fashion. We have found that the damped oscillation is well approximated by
\begin{equation}
\delta G_+^{CFT}(t,t;k)\approx\Big(\!A_1 \cos (2k t\!+\!\alpha_1)\!+\!B_1\!\Big) e^{-4\pi t T_f \Delta},\label{eq:quasi}
\end{equation}
where $(A_1,B_1,\alpha_1)$ are constants that depend on the value of $k$. The red dashed curve in Fig.\ \ref{fig:Decayfig} shows a fit of the form (\ref{eq:quasi}) together with the numerically calculated boundary theory Wightman function (blue solid curve), with $(A_1,B_1,\alpha_1)$ chosen in order to obtain a good fit at late times. The exponential decay rate in (\ref{eq:quasi}) is equivalent to twice the imaginary part of the lowest BTZ quasinormal mode, $\omega_i=\Im(\omega_*^{(0)})=-2\pi T_f\Delta$. In addition, the oscillation frequency in (\ref{eq:quasi}) is twice the real part of the lowest quasinormal mode, $\omega_r=\textrm{Re}(\omega_*^{(0)})=k$.

\section{An argument for the quasinormal decay in AdS-Vaidya}\label{sec:3}

In this section, we present an argument for the appearance of the quasinormal decay of the Wightman function (\ref{eq:quasi}) that we observe in the numerical calculation. The argument consists of two steps
presented in the subsections \ref{sec:argument1} and \ref{sec:argument2}.
First, in Section \ref{sec:argument1}, we remind the reader that smooth scalar field configurations in AdS-Vaidya spacetime are expected to decay for $v\!>\!0$ with a rate set by the lowest quasinormal mode of the black hole. For the case of AdS$_3$-Vaidya, and $\Delta = 3/2$, we show this explicitly, while leaving many of the details to Appendices. We point out which parts of this argument apply to higher dimensions and which parts lack a generalization to higher dimensions. In higher-dimensional cases, the quasinormal decay of smooth initial data is still expected, based on numerical simulations (see e.g.\ \cite{Heller:2013oxa}). As a second part of the argument, in Section \ref{sec:argument2}, we assume the quasinormal decay of smooth initial data and show how this implies the observed quasinormal decay of the Wightman function.

\subsection{Quasinormal decay of smooth scalar field configurations}\label{sec:argument1}

Consider solving the Klein-Gordon equation of motion $(\Box-m^2)\phi=0$ in the AdS-Vaidya background. Under a Fourier transform with respect to the boundary spatial coordinates, the scalar field becomes a function of the variables $(v,z,k)$. Given the profile $\phi(v\!=\!0,z,k)=\phi_0(z,k)$ of the solution at time $v\!=\!0$, the solution at a later time $v\!>\!0$ is given by
\begin{equation}
	\phi(v,z,k)=i\!\int_{v'=0}\!\!\!\!\!\!\!\! dz'\, \phi_0(z',k)\overleftrightarrow{D}^{\!v'}G_R(v,z;v'\!,z'\!;k),\label{eq:generalsol}
\end{equation}
where $G_R(v,z;v',z';k)$ is the retarded propagator in the static black hole background and $D^v=\sqrt{-g}g^{vz}\partial_z$. The reason why we can use the retarded propagator of the static black hole is that the retarded propagator in the $v\!\ge\! 0$ region does not depend on the form of the spacetime for $v\!<\!0$. We refer the reader to \cite{Keranen:2014lna} for a detailed explanation of this point.

Since we are interested in boundary theory quantities in the end, we consider the boundary limit $\phi^b(v,k)=\sqrt{2\pi}\lim_{z\to 0}z^{-\Delta}\phi(v,z,k)$. In this limit, (\ref{eq:generalsol}) becomes
\beq
	\phi^b(v,k) = i \!\int_{v'=0}\!\!\!\!\!\!\!\! dz'\, \phi_0(z',k)\overleftrightarrow{D}^{\!v'}G^{bb}_R(v;v',z'\!;k),\label{eq:generalsol2}
\eeq
where $G_R^{bb}$ denotes the bulk-to-boundary propagator
\begin{equation}
	G_R^{bb}(v;v'\!,z'\!;k)=\!\sqrt{2\pi}\lim_{z\to 0}z^{-\Delta}G_R(v,z;v'\!,z'\!;k).
	\label{eq:bulk_to_boundary}
\end{equation}
When the two points $(0,v)$ and $(z',v')$ have a large timelike separation, the retarded propagator decays exponentially according to the quasinormal modes $\omega_*^{(n)}$ of the black hole background,
\beq
	G^{bb}_R(v;v',z';k)\approx\sum_n c_n(z') e^{-i\omega_*^{(n)}(v-v')}.\label{eq:Grdecay}
\eeq
This follows from a definition of the quasinormal modes as the poles of the retarded propagator in Fourier space. 
Furthermore, the leading term in (\ref{eq:Grdecay}) at late times is given by the lowest quasinormal mode
\beq
	G^{bb}_R(v;v',z';k)\approx c_0(z') e^{-i\omega_*^{(0)}(v-v')}+c.c.,\label{eq:Grdecay2}
\eeq
where the complex conjugate term follows from the fact that the lowest quasinormal mode is degenerate with the quasinormal mode $-(\omega_*^{(0)})^*$.

For the case of the AdS$_3$-Vaidya spacetime, $G_R^{bb}$ is the BTZ retarded propagator, which
is analytically known for a $\Delta =3/2$ scalar field and given in (\ref{eq:GRbb}). The expansion in quasinormal modes (\ref{eq:Grdecay}) in this case is studied in Appendix~\ref{sec:QNM} and the first few terms explicitly spelled out in (\ref{eq:Grbb_quasinormal}). For example, the lowest quasinormal mode contribution is given by
\begin{align}
	 iG_R^{bb}&(v;v',z';k)\notag\\
		\approx&\left(\!\!\frac{(4\pi T_fz')^{\frac{3}{2}}\Gamma\left(\!-\frac{ik}{2\pi T_f}\!\right)e^{-i k(v-v')}}{(1\!-\!2\pi T_f z')^{\frac{3}{2}+\frac{ik}{2\pi T_f}}\Gamma\left(\!-\frac{1}{2}\!-\!\frac{ik}{2\pi T_f}\!\right)}+c.c.\!\!\right)\times\notag\\
		&\qquad\qquad\times e^{-\frac{3}{2} 2\pi T_f (v-v')}.
	\label{eq:approx2text}
\end{align}
A shortcoming of the quasinormal mode expansion (\ref{eq:Grdecay}) is that it is not valid when the points $(0, v)$ and $(z', v')$ are lightlike separated. We will denote the point where this happens as $z' = z^*_a$.  At late times, $z^*_a$ is near the black hole horizon $z^*_a \approx 1/(2\pi T_f)$. The way we will proceed is to split the $z'$ integral (\ref{eq:generalsol2}) into a region near the lightcone $z'\in (z^*_a -\zeta, z^*_a)$ and a region away from the lightcone $z'\in (0, z^*_a -\zeta)$, where $\zeta$ is taken to be a small positive real number,
\begin{align}
	\phi^b(v,k) =\,& i\!\! \int_{0}^{z_a^*-\zeta}\!\!\!\!\!\!\!\! dz' \phi_0(z',k)\!\overleftrightarrow{D}^{\!v'}\!\!G^{bb}_R(v;v'\!\!=\!0,z'\!;k)\notag\\
&\!\!+\! i \!\!\int_{z_a^*-\zeta}^{z_a^*}\!\!\!\!\!\!\! dz' \phi_0(z',k)\!\overleftrightarrow{D}^{\!v'}\!\!G^{bb}_R(v;v'\!\!=\!0,z'\!;k)
.\label{eq:generalsol3}
\end{align}
In the first line of (\ref{eq:generalsol3}), the integral covers the region away from the lightcone singularity, in which case we can use the quasinormal mode expansion inside the integral. Thus, the first integral in (\ref{eq:generalsol3}) decays at late times with a rate set by the lowest quasinormal mode. However, it is not obvious how the second line of (\ref{eq:generalsol3}) behaves at late times. In Appendix~\ref{sec:integral}, we show that for AdS$_3$-Vaidya spacetime, and $\Delta = 3/2$, the integral near the lightcone can be analytically computed and is shown to decay with a rate set by the  lowest quasinormal mode, as long as the initial data $\phi_0(z',k)$ is a smooth function of $z'$ near the black hole horizon. Thus, we are led to the late time behavior
\beq
	\phi^b(v,k)\approx a e^{-i(\omega_r+i\omega_i)v}+ b e^{-i(-\omega_r+i\omega_i)v},\label{eq:decay1}
\eeq
where we denote the pair of lowest quasinormal modes as $\omega_*^{(0)}=\pm \omega_r+i\omega_i$. 

Showing that the integral near the lightcone singularity in (\ref{eq:generalsol3}) decays in time with a rate set by the lowest quasinormal mode also in the case of higher dimensions is out of the scope of this paper. The most straightforward way to convince oneself that (\ref{eq:decay1}) is still true more generally is to solve the equation of motion of the scalar field numerically.  For example, for a massless scalar in the AdS$_5$-Schwarzschild spacetime such a numerical calculation confirming the quasinormal decay of smooth initial data can be found in \cite{Heller:2013oxa}.\footnote{The relevant calcuation in \cite{Heller:2013oxa} is phrased as solving the linearized Einstein's equations in a certain channel, but the resulting equation can be shown to be equivalent to the massless Klein-Gordon equation.} In what follows, we assume that the quasinormal decay (\ref{eq:decay1}) holds for smooth initial data. Thus, the rest of this section can be applied to any AdS$_d$-Vaidya spacetime, where this assumption holds. In particular, we have shown that it holds for the AdS$_3$-Vaidya case for $\Delta = 3/2$.

\begin{figure}[t]
\begin{center}
\includegraphics[width=0.975\linewidth]{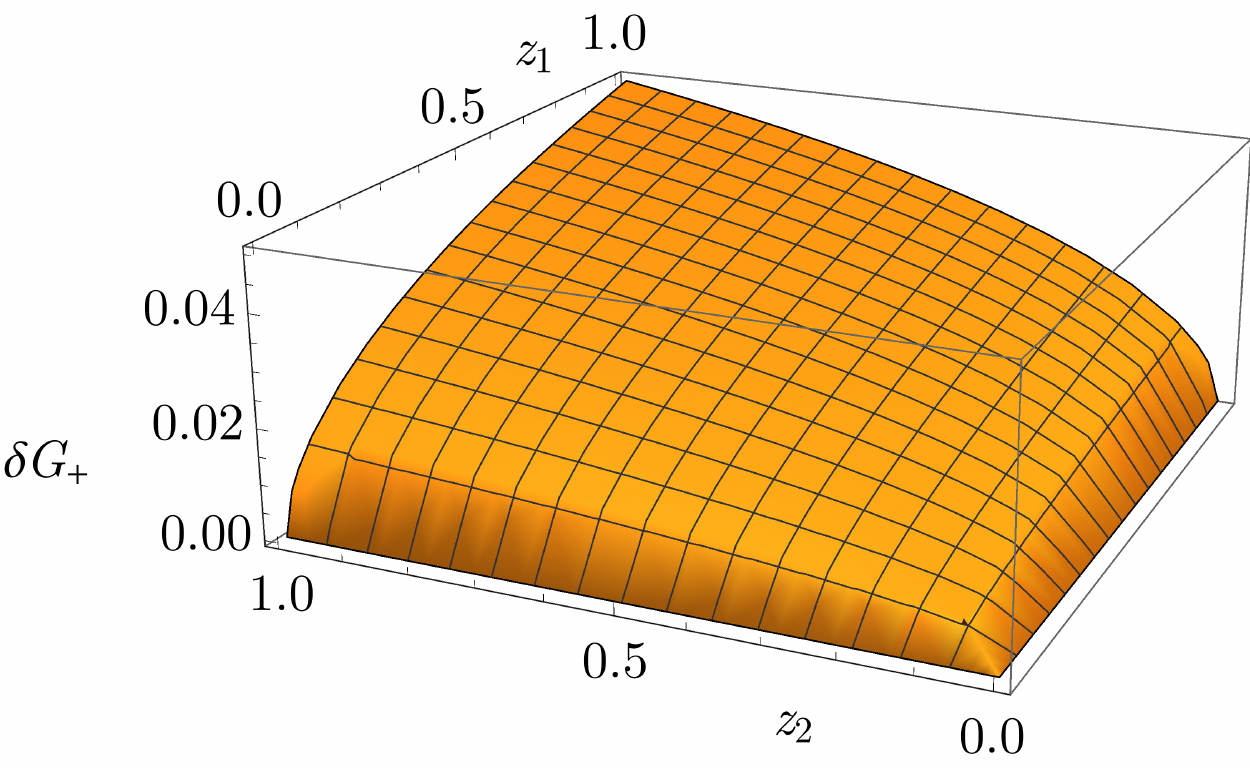}
\caption{\label{fig:initialdata} 
Initial data on the $v\!=\!0$ hypersurface for the subtracted bulk Wightman function (\ref{eq:differencecorr}) in the case of AdS$_3$-Vaidya. We have chosen units such that the horizon is located at $z\!=\!1$ and the value of momentum is given by $k\!=\!1$, which we have found to be a good representative of the situation for general $k$. Unlike the Wightman functions themselves, their difference is a smooth function in the bulk.}
\end{center}
\end{figure}

\subsection{Quasinormal decay of the Wightman function}\label{sec:argument2}

Consider the quantity
\begin{align}
\delta G_+(v_1,z_1;v_2,z_2;k)
=&\;G_+(v_1,z_1;v_2,z_2;k)\nonumber
\\
&-G_+^{th}(v_1,z_1;v_2,z_2;k)\label{eq:differencecorr}
\end{align}
in the Vaidya spacetime. Here $G_+^{th}$ denotes the black hole Wightman function in the Hartle-Hawking state, which, in the case of the BTZ black hole, is given in (\ref{eq:GfBTZ}) in Appendix~\ref{sec:method_old}. For $v_1\!\ge\! 0$ and $v_2\!\ge\! 0$, the quantity (\ref{eq:differencecorr}) clearly satisfies the Klein-Gordon equation with respect to both of its arguments. Furthermore, at $v_1\!=\!v_2\!=\!0$, $\delta G_+$ is a smooth function of $z_1$ and $z_2$, thus providing smooth initial data for the time evolution. Note that while both $G_+^{AdS}$ and $G_+^{th}$ have singularities at lightlike separation, their difference does not. This is because in a quantum field theory, the leading short distance singularities of correlation functions are independent of the state for any reasonable state. Alternatively, one can check this explicitly in AdS$_3$ where the functions are known analytically. For concreteness, we plot the initial data for AdS$_3$-Vaidya and $\Delta=3/2$ in Fig.\ \ref{fig:initialdata} where one can observe that it is smooth as a function of $z_1$ and $z_2$. 

In order to obtain the boundary Wightman function (\ref{eq:boundary_correlator})
in the black hole region, we can time evolve $\delta G_+(v\!=\!0,z;v'\!=\!0,z';k)$ applying (\ref{eq:decay1}) as it provides smooth initial data on the hypersurface of the infalling shell. We can first time evolve $(v\!=\!0,z)$ to $(v_1,z_1)$ to get 
\begin{align}
	\lim_{z_1\to0} z_1^{-\Delta}&\delta G_+(v_1,z_1;v'\!=\!0,z';k)\notag\\
	&\propto  f(z') e^{(\omega_{i}-i\omega_{r})v_1-ig(z')}+c.c.,\label{eq:12}
\end{align}
where $f(z')$ and $g(z')$ are some real valued smooth functions of $z'$, and we have taken the limit $z_1\rightarrow 0$ anticipating the fact that we want to obtain the boundary theory correlator in the end. We can take (\ref{eq:12}) as initial data to be evolved forwards in time from $(v'\!=\!0,z')$ to $(v_2, z_2)$, which results in
\begin{align}
	\lim_{z_1,z_2\to0}& (z_1 z_2)^{-\Delta}\delta G_+(v_1,z_1;v_2,z_2;k)\label{eq:long}\\
		\propto\, & e^{(\omega_i-i\omega_r)v_1}\Big(A_2 e^{(\omega_i-i\omega_r) v_2+i\alpha_2}\notag\\
		&\hspace{2cm}+B_2 e^{(\omega_i+i\omega_r) v_2+i\gamma_2}\Big)+c.c.\notag\\
		=\,&2 e^{\omega_i(v_1+v_2)}\Big(A_2\cos(\omega_r(v_1+v_2)-\alpha_2)\notag\\
		&\hspace{2cm}+B_2\cos(\omega_r(v_1-v_2)+\gamma_2)\Big),\notag
\end{align}
where $(A_2,B_2,\alpha_2,\gamma_2)$ are constants that depend on the momentum. Thus, we are lead to conclude that $\delta G_+^{CFT}$ decays with the lowest quasinormal mode in the way we observed from the numerical calculation in (\ref{eq:quasi}).

\section{A simpler numerical method for obtaining the Wightman functions}\label{sec:4}

The previous discussion suggests a simpler method for calculating the AdS-Vaidya Wightman functions than the method used in \cite{Keranen:2014lna} while also providing a concrete check of the above arguments. Rather than solving the equations of motion (\ref{eq:eoms}) for the Wightman function itself, we solve them for the difference $\delta G_+(v_2,z_2;v_1,z_1;k)$. Solving the equations of motion with respect to both of the arguments is now straightforward numerically as the initial data $\delta G_+(0,z_1;0,z_2;k)$ is smooth. We have used the method of Green's functions to solve for the two-point function
\begin{align}
	\delta G_+(v_2,z_2;v_1,z_1;k)&\label{eq:integral_1}\\
		=-\int dz dz' \,& \Big[\delta G_+(0,z';0,z;k)\times\notag\\
		&\times\! \overleftrightarrow{D}^{v'} G_R^{BTZ}(v_2,z_2;0,z';k)\times\notag\\
		&\times\! \overleftrightarrow{D}^{v} G_R^{BTZ}(v_1,z_1;0,z;k)\Big].\notag
\end{align}
We refer to Appendix \ref{sec:method_new} for more details on the method. We have checked that, up to numerical accuracy, the new method of calculating the Wightman function agrees with our previous one while being numerically considerably simpler.

\section{Effective occupation numbers}\label{sec:5}

In this section, we change gears and consider some concrete information that can be extracted from the Wightman functions, namely the occupation numbers in the boundary field theory. A strategy for obtaining an effective occupation number for nonequilibrium systems was presented in the context of holography in \cite{Balasubramanian:2012tu}, inspired by the fluctuation dissipation relation in thermal equilibrium. The same definition was earlier used in nonequilibrium quantum field theory \cite{Berges:2010ez}. Here, we will calculate the effective occupation number in the system at hand, the 1+1 dimensional conformal field theory dual to the AdS$_3$-Vaidya collapse. The effective occupation number is defined as\footnote{We have defined the  retarded two-point function following the convention of \cite{Keranen:2014lna}, which also leads to the unfamiliar looking relation, that the spectral function is given by the real part of the retarded correlator as opposed to the usual convention in which it is the imaginary part.}
\beq
	n_{eff}(\omega,k,t)=\frac{G_-(\omega,k,t)}{2\textrm{Re}\, G_R(\omega,k,t)},\label{eq:n}
\eeq
where we have performed a Wigner transform of the Wightman and the retarded two-point function,
\beq
	G(\omega,k,t)=\int\!\! d\delta t\, e^{i\omega\delta t}G(t\!+\!\delta t/2,t\!-\!\delta t/2;k)\label{eq:wigner}.
\eeq
Due to the fluctuation dissipation relation, the effective occupation number (\ref{eq:n}) reduces to the thermal distribution
\beq
	n_{BE}=\frac{1}{e^{\beta \omega}-1}
\eeq
in thermal equilibrium. Thus, it provides a fairly natural generalization of the equilibrium occupation number. We calculate the effective occupation number by first computing $G(t\!+\!\delta t/2,t\!-\!\delta t/2;k)$ for a range of discretized values of $t$ and $\delta t$ and then numerically performing the Fourier transform with respect to $\delta t$.

\begin{figure}[t]
\begin{center}
\includegraphics[width=0.975\linewidth]{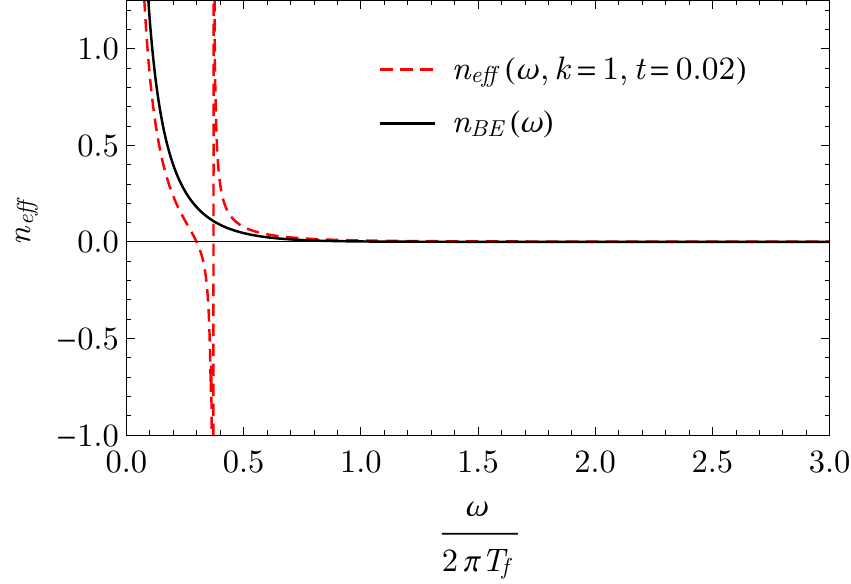}
\caption{\label{fig:n1} 
The red dashed curve is the effective occupation number $n_{eff}$ for average time $t\!=\!0.02/(2\pi T_f)$ as a function of energy $\omega$ and at momentum $k=2\pi T_f$, while the black solid curve is the Bose-Einstein distribution corresponding to the final temperature $T_f$.}
\end{center}
\end{figure}

As an example, we show the effective occupation number at a fixed average time $t\!=\!0.02/(2\pi T_f)$ in Fig.\ \ref{fig:n1}.  The first thing to note is that the effective occupation number differs from the thermal one. This is not very surprising as we are considering a nonequilibrium quench state in the CFT. Secondly, the occupation number has poles as a function of $\omega$. This is not quite as surprising as it might seem at first. Even in thermal equilibrium $n_{BE}$ has a pole at $\omega\!=\!0$. The quantity that is relevant for counting the number of occupied states is the combination $\Omega(\omega)n_{BE}(\omega)$, where $\Omega(\omega)$ is the density of states. For a free boson quantum field theory in thermal equilibrium, one can show that the number of occupied particle states with energy less than $E$ is given by the integral
\beq
N(E)=\!\int\!\! d^{d-1}x\! \int_{k^0\ge 0}^{k^0\le E}\!\!\!\! \frac{d^dk}{(2\pi)^d}\,2\omega \rho(\omega,k)n_{BE}(\omega),\,\,
\eeq
where $\rho\!=\!2\,\textrm{Re}(G_R(\omega,k))$ is the thermal spectral function\footnote{See previous footnote.} of the free boson. This example suggests that the quantity that appears in counting occupied states is the combination $\rho\, n$. In the following, we will therefore plot the combination $\rho(\omega,k,t) n_{eff}(\omega,k,t)$, where $\rho$ is the nonequilibrium spectral function. The time evolution of $\rho$ has been studied before in \cite{Balasubramanian:2012tu}. The combination $\rho\, n_{eff}$ has no poles, as all the poles in $n_{eff}$ coincide with the zeros of $\rho$. Thus, the positions of the poles in $n_{eff}$ correspond to energies where the effective density of states vanishes. 

\begin{figure}[t]
\begin{center}
\includegraphics[width=0.975\linewidth]{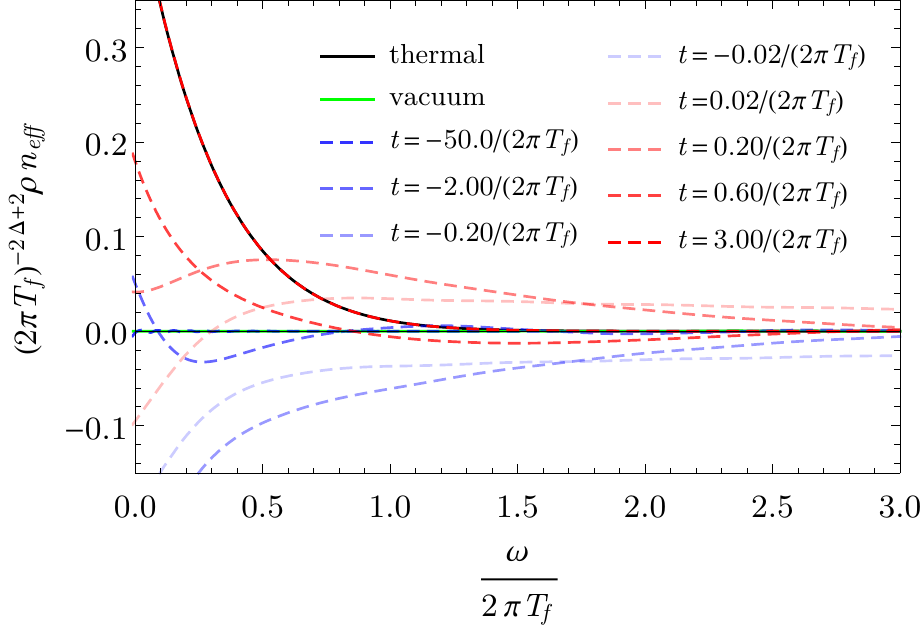}
\caption{\label{fig:n2} 
The combination $\rho(\omega,k,t) n_{eff}(\omega,k,t)$, which counts occupied states, is plotted as a function of $\omega$ for different times at momentum $k=2\pi T_f$. Initially, at large negative times, no states are occupied and $\rho\, n_{eff}$ vanishes. Around the time of the quench, $\rho\, n_{eff}$ oscillates until it finally settles to its thermal value.}
\end{center}
\end{figure}

In Fig.\ \ref{fig:n2}, $\rho\, n_{eff}$ is plotted for different values of time as a function of $\omega$. At very early times, it starts from being very close to zero. This is indeed expected as the field theory is prepared in the vacuum state, so no states should be occupied. As time progresses, it first oscillates to a negative value, then later becomes positive. At the intermediate times near $t\!=\!0$, $\rho\, n_{eff}$ is smaller than the thermal value (black curve) for small $\omega$, while it is larger than the thermal value for large $\omega$. This suggests that after the quench, the low energy states are underoccupied 
and the high energy states overoccupied, as compared to thermal equilibrium. In the regions where it is negative, we do not expect it to have an interpretation in terms of an actual occupation number.

\begin{figure}[t]
\begin{center}
\includegraphics[width=0.975\linewidth]{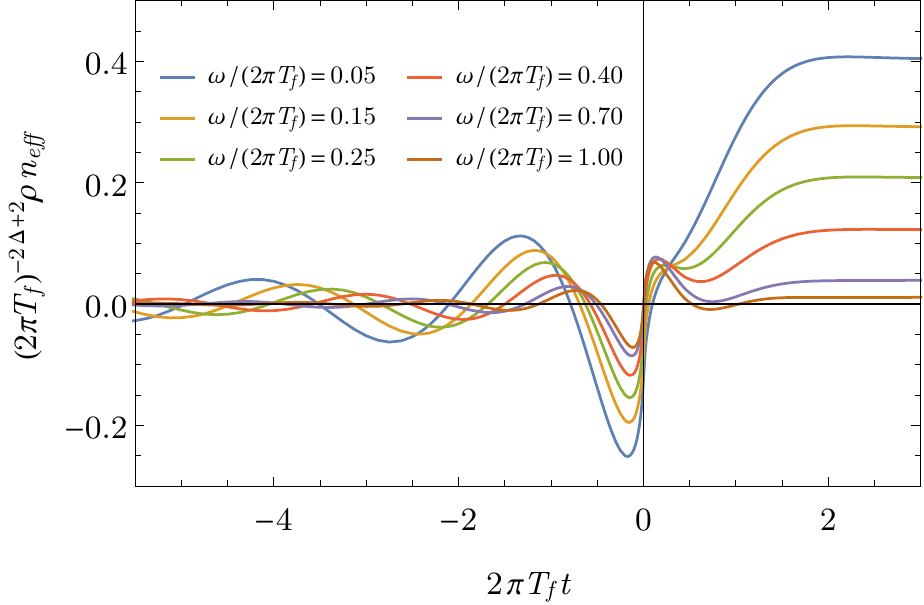}
\caption{\label{fig:n3} 
The quantity $\rho(\omega,k,t)n_{eff}(\omega,k,t)$, which counts occupied states, is plotted as a function of $t$ for different values of $\omega$ and fixed $k\!=\!2\pi T_f$. $\rho\, n_{eff}$ starts oscillating well before the quench at $t\!=\!0$ as it is sensitive to all times due to the Wigner transform in (\ref{eq:wigner}). The largest changes happen at the time of the quench before the quantity settles down to its thermal value with $n_{eff}\to n_{BE}$ at late times.}
\end{center}
\end{figure}

\begin{figure}[t]
\begin{center}
\includegraphics[width=0.975\linewidth]{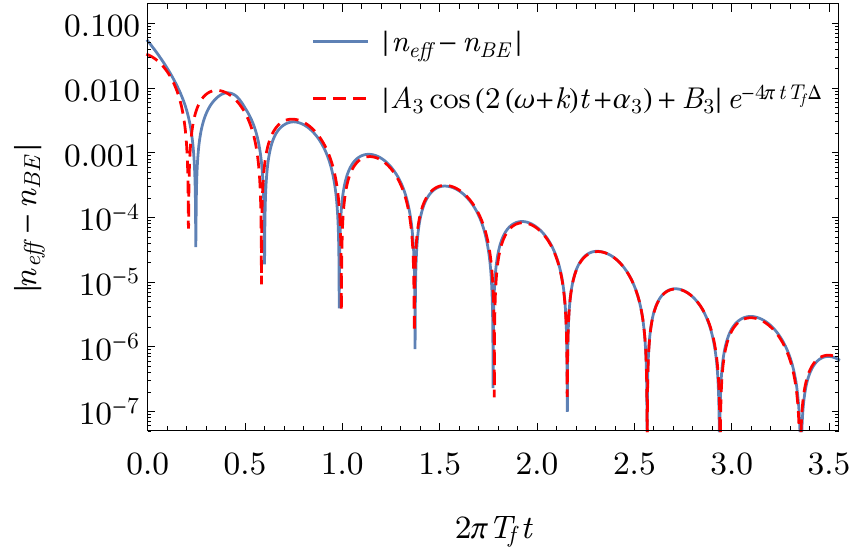}
\caption{\label{fig:n4} 
The absolute difference between the effective occupation number $n_{eff}(\omega,k,t)$ and the thermal Bose-Einstein distribution $n_{BE}(\omega)$ is plotted as a function of $t$ with $k\!=\!4\pi T_f$ and $\omega\!=\!4\pi T_f$. At late times, the difference shrinks exponentially with the time scale set by the lowest quasinormal mode of the black hole background.}
\end{center}
\end{figure}

Fig.\ \ref{fig:n3} shows $\rho\, n_{eff}$ as a function of time for fixed values of $\omega$ and $k$. This shows the same basic features as Fig.\ \ref{fig:n2}. At late times, $\rho\, n_{eff}$ approaches a constant value which coincides with the thermal value. Fig.\ \ref{fig:n4} shows a plot of the difference of the effective occupation number and the thermal Bose-Einstein distribution. The effective occupation number decays towards the thermal distribution in a form that is well approximated by
\beq
n_{eff}\!-\!n_{BE}\!\approx\! e^{-4\pi t T_f \Delta}[A_3 \cos(2(\omega\!+\!k)t\!+\!\alpha_3)\!+\!B_3]\,\,\,\,\,\,
\eeq
at late times, where $(A_3,B_3,\alpha_3)$ are constants, which depend on the frequency and momentum. This is the same exponential rate at which the equal-time Wightman function decays towards equilibrium, the lowest quasinormal mode of the scalar field. In contrast, the oscillation frequency of $n_{eff}\!-\!n_{BE}$ differs from the real part of the lowest quasinormal mode due to the Wigner transform with respect to relative time.

As discussed in \cite{Balasubramanian:2012tu}, it can be useful to consider the effective occupation number, where, when performing the Wigner transform, one inserts a smearing function inside the $\delta t$ integral in (\ref{eq:wigner}) (for example a Gaussian centered around $\delta t\!=\!0$). This suppresses the contributions of large time separation $\delta t$ to the Wigner transform and makes the effective occupation number a quantity more local in time. We have done this for a Gaussian window function in Appendix \ref{sec:a}, where the reader can find more details on this procedure. The main differences to the results shown in the main text are that the early time fluctuations of $\rho\, n_{eff}$ are not present there. However, the early and late time limits of the effective occupation numbers are no longer given by $0$ and the Bose-Einstein distribution, respectively.

\section{The geodesic approximation in AdS$_3$-Vaidya}\label{sec:6}

\begin{figure}[t]
\begin{center}
\includegraphics[width=0.975\linewidth]{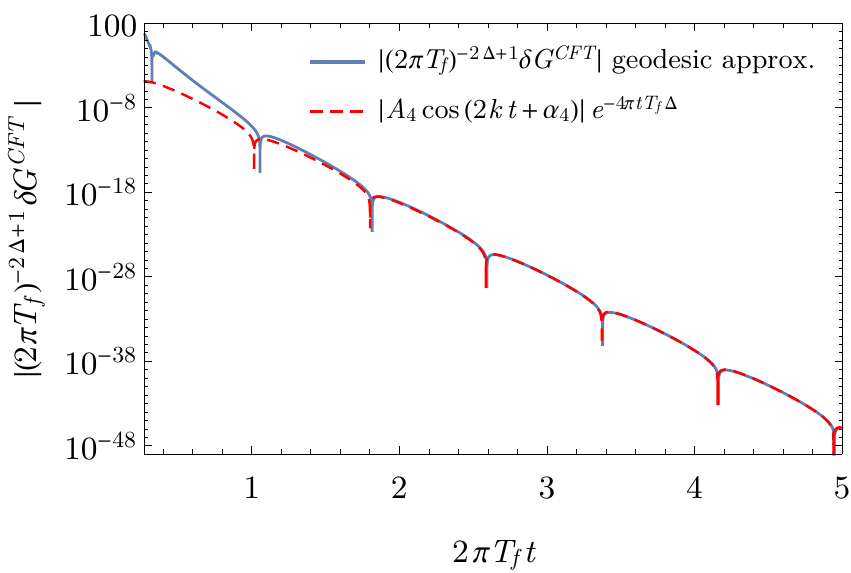}
\caption{\label{fig:geod} 
The Fourier transformed correlator in the geodesic approximation for $\Delta=10$, $k=4\pi T_f$ and $\delta t=0.01/(2\pi T_f)$. The red dashed curve is a fit to the lowest BTZ quasinormal mode.}
\end{center}
\end{figure}

We would like to ask whether one obtains the result (\ref{eq:quasi}) for large $\Delta$ using the geodesic approximation. It is not completely obvious that the geodesic approximation can be trusted in the context of AdS-Vaidya spacetime. Firstly, it is not clear which inverse of the operator $\Box-m^2$ the geodesic approximation is computing, since the inverse is no longer unique in a real-time situation. Secondly, there is an assumption on analytic continuation that must be made to rotate the particle path integral to pass through the saddle point. It is not clear whether this analytic continuation is justified in AdS-Vaidya spacetime, as the metric is not an analytic function of time. For more details on these problems, we refer the reader to \cite{Louko:2000tp}. With these cautionary points in mind, we will make use of the geodesic approximation in the following. The CFT two-point function is obtained from the (renormalized) geodesic length $L_{ren}(t_1,x_1;t_2,x_2)$ of the minimal length bulk geodesic connecting the boundary points $(t_1,x_1)$ and $(t_2,x_2)$. The Fourier transform of the subtracted correlator is
\begin{align}
&\delta G^{CFT}(t_1,t_2;k)\label{eq:geod}
\\
&\!\!=\!\int_{-\infty}^{\infty}\!\!\!\! dk\, e^{ikx}\Big(e^{-\Delta L_{ren}(t_1,x;t_2,0)}\!-\!e^{-\Delta L_{ren}^{BTZ}(t_1,x;t_2,0)}\Big).\nonumber
\end{align}
The procedure for calculating the geodesic can be found in \cite{AbajoArrastia:2010yt,Albash:2010mv,Balasubramanian:2011ur} and the relevant formulas are compactly collected in an Appendix of \cite{Keranen:2014lna}. Here, we will simply give the results of the numerical evaluation of (\ref{eq:geod}). In order to evaluate the Fourier transform, we need to specify a value for $\Delta$. In Fig.\ \ref{fig:geod}, we have chosen $\Delta=10$ and $k=4\pi T_f$ to evaluate the integral. We have found that different values of $\Delta$ and $k$ lead to similar results. As shown in Fig.\ \ref{fig:geod}, the Fourier transformed geodesic correlator decays at late times to a good approximation with a rate and oscillation frequency set by the lowest quasinormal mode.

\section{Conclusions}\label{sec:7}

In AdS$_3$-Vaidya spacetime, we have presented numerical and analytical evidence that the boundary theory Wightman two-point functions thermalize with the rate set by the lowest quasinormal mode of the corresponding bulk field in the BTZ background. This is not too surprising from the point of view of the bulk problem as the quasinormal modes appear as solutions of the same equation of motion as the bulk Wightman function satisfies. A crucial observation is that the difference between the thermal BTZ bulk two-point function and the bulk two-point function in the Vaidya spacetime satisfies the Klein-Gordon equation for $v\!\geq\!0$ while it has no singularities. Therefore, one can expect it to decay at later times as any smooth scalar perturbation of the final black hole, i.e.\ with the lowest quasinormal mode. This argument is expected
to hold also in higher-dimensional AdS$_d$-Vaidya spacetime, and for arbitrary values of the mass $m$. 
Other examples of holographic settings where the two-point function, starting far from thermal equilibrium, relaxes towards equilibrium according to the lowest quasinormal mode are studied in \cite{Hartman:2013qma,Guica:2014dfa}.

The big picture that has emerged from the study of nonequilibrium one-point functions in holography is that, as long as there is black hole formation in the bulk, the subsequent dynamics of the one-point functions is controlled by the lowest quasinormal mode (see e.g.\ \cite{Bhaseen:2012gg,Craps:2015upq}). Our results suggest that the same picture applies to two point correlation functions. In the presence of a horizon, they decay towards equilibrium with a rate set by the lowest quasinormal mode. Often, nonequilibrium two-point functions are calculated in the geodesic approximation, in which case this behavior is not completely obvious. In section \ref{sec:6} however, we found numerical evidence that one indeed also finds the quasinormal decay in the geodesic approximation after performing the Fourier transform to momentum space. 

The quasinormal decay we find in Wightman functions should be contrasted with what one finds for the boundary retarded correlation function \cite{Balasubramanian:2012tu,Callebaut:2014tva,David:2015xqa}. The retarded correlation function $G_R(t_1,t_2;k)$ decays exponentially, approximately with the lowest quasinormal mode, when the later time argument $t_1$ is taken to the black hole region $t_1\!>\!0$. On the other hand, when both of the points in the retarded correlator are taken to the black hole region, $t_1\!>\!0$ and $t_2\!>\!0$, the retarded correlator in the Vaidya spacetime becomes immediately thermal. This is not what happens for the Wightman function, which instead approaches its thermal value exponentially in the time variable $t_1\!+\!t_2$.

Finally, we would like to comment on the general thermalization pattern in holography in a more speculative vein. In the study of two-point functions in the geodesic approximation (and Wilson loops and entanglement entropy), it has been found that the correlators with short spacelike separation thermalize earlier than the correlators with large spacelike separation. As we have shown, the momentum space two-point functions decay towards thermal equilibrium with a rate that is independent of momentum in the AdS$_3$-Vaidya example. Naively, one might have expected that the faster thermalization of the short distance correlators would imply faster thermalization of the large momentum correlators. The AdS$_3$-Vaidya example we have studied in this paper shows that this expectation is incorrect. Here, all momenta thermalize on the same time scale $1/(\pi T_f \Delta)$ set by the lowest quasinormal mode, which in the AdS$_3$ case is independent of momentum.\footnote{Since the exponential decay is independent of the momentum in this case, it follows that the overall value of the large momentum correlators is at all times closer to the thermal correlator than for small momentum, because this is the case already in the initial vacuum state. On the other hand, the ratio $\delta G_+(t+\delta t/2,t-\delta t/2;k)/\delta G_+(\delta t/2,-\delta t/2;k)$, which quantifies the deviation from thermality as compared relative to the initial deviation from thermality, does not become smaller as $k$ is increased.}
Furthermore, in \cite{Chesler:2012zk}, it was shown that for momentum space two-point functions in an AdS$_5$ nonequilibrium black hole, large momenta and frequencies thermalize the slowest. This is the opposite of what one would expect from combining results of the geodesic approximation with the naive expectation that large momentum corresponds to short distances. Our picture of the thermalization rate being dictated by the lowest quasinormal mode is consistent with the results of \cite{Chesler:2012zk} since, for AdS$_5$-Schwarzschild black holes, the imaginary part of the lowest scalar quasinormal mode approaches zero as $k^{-1/3}$, signaling slow thermalization of the large momentum correlator. We believe that the apparent difference from the geodesic approximation results could follow from the incorrect expectation that large momentum always corresponds to short distance in the correlation function.\footnote{In a Fourier transform of a function $f(x)$, the high momenta are of course sensitive to the short-scale features of the function $f(x)$. But when one applies this to the two-point function $\int dx\, e^{ikx}\langle\mathcal{O}(x,t_1)\mathcal{O}(0,t_2)\rangle$, the relevant short-scale features need not be located near $x=0$, when there are many scales in the problem. Indeed, from the geodesic approximation one learns that the function $\langle\mathcal{O}(x,t_1)\mathcal{O}(0,t_2)\rangle$ has sharp features at $x\propto t_1+t_2$, which at late times is certainly not at short distances but can give a sizeable contribution to the
large momentum Fourier transform.} We leave a detailed study of this problem for future work.

\section{Acknowledgments}

We would like to thank Jorge Casalderrey-Solana, Esko Keski-Vakkuri, Andrei Starinets and Larus Thorlacius for useful discussions. This research was supported by the European Research Council under the European Union's Seventh Framework Programme (ERC Grant Agreement 307955).

{\parindent0pt \parskip10pt \textit{Note added.}---While this paper was being finished, \cite{Lin:2015acg} appeared, which has some overlap with our work, in particular our Section\ \ref{sec:4}.}

\appendix

\section{``Old'' method for computing the out-of-equilibrium Wightman function}\label{sec:method_old}

The out-of-equilibrium Wightman functions can be obtained by solving the Klein-Gordon equations of motion (\ref{eq:eoms}) given the Wightman function at early times in the initial state of interest. In this appendix, we review details of the methods from \cite{Keranen:2014lna}, which we use to solve (\ref{eq:eoms}).

The initial state we are interested in is the AdS$_3$ ground state for the scalar field. The bulk-to-bulk Wightman function in AdS$_3$, for our choice of scaling dimension $\Delta = 3/2$, is given by
\begin{align}
	 G_{+}^{\text{AdS}}&(v_2,z_2;v_1,z_1;k)\label{eq:GfAdS}\\
		=&\, \frac{\sqrt{z_1 z_2}}{2\pi}\left[K_0\left(\sqrt{\sigma}\,|k|\right)-K_0\left(\sqrt{\sigma+4z_1 z_2}\,|k|\right)\right],\notag
\end{align}
where $K_0$ is a modified Bessel function and
\begin{equation}
	\!\!\!\sigma\!=\!-(v_2\!-\!v_1)^2\!-2(v_2\!-\!v_1)(z_2\!-\!z_1)+(v_2\!-\!v_1\!+\!z_2\!-\!z_1)i\epsilon.
\end{equation}
While \cite{Keranen:2014lna} phrased all the calculations in terms of the Feynman correlator, here we will consider the Wightman correlator, in which case the $i\epsilon$ factors are different. The Wightman correlator can be obtained from the Feynman correlator using the identity
\begin{align}
G_+&(x_2,x_1)\\
&=\theta(t_2\!-\!t_1)G_F(x_2,x_1)+\theta(t_1\!-\!t_2)G_F^*(x_2,x_1),\notag
\end{align}
where $x$ denotes all the coordinates in question and $t$ is the time coordinate. Our task is to solve (\ref{eq:eoms}) with the initial data that the correlator in the early time (that is, for $v_1<0 $ and $v_2<0$) is given by (\ref{eq:GfAdS}). This uniquely specifies the correlation function at later times. We use the method of Green's functions to solve for the time evolution with respect to the time coordinate $v_2$,
\begin{align}
G_+&(v_3,z_3;v_1,z_1;k)\label{eq:jformula_ini}\\
	=\,& i\!\!\int_{v_2=0}\!\!\!\!\!\!\!\!\!\! dz_2\Big[G_{\!+}(v_2,z_2;v_1,z_1;k)\!\overleftrightarrow{D}^{\!v_2}G_{\!R}(v_3,z_3;v_2,z_2;k)\Big],\notag
\end{align}
where $G_R(v_3,z_3;v_2,z_2;k)$ is the retarded correlation function and we have defined the differential operator $D^v=\sqrt{-g}g^{v\mu}\partial_{\mu}$. The reason the integral form (\ref{eq:jformula_ini}) simplifies the calculation is as follows. The retarded propagator depends only on the spacetime in the future light-cone of the point $(v_2, z_2)$. This is a property of free fields and is reviewed for example in \cite{Keranen:2014lna}. Since we have chosen this point to lie on the surface $v_2=0$, where the AdS and BTZ parts of the Vaidya spacetime are joined, the retarded propagator is identical to the retarded propagator in the BTZ spacetime, which is analytically known. We can extract the BTZ retarded propagator from the BTZ Wightman function via
\begin{equation}
	G_R^{BTZ}(x_2,x_1)=2i\theta(t_2-t_1)\Im G_+^{BTZ}(x_2,x_1).
\end{equation}
For $T_f=1/2\pi$, the BTZ Wightman two-point function is given by
\begin{widetext}
\vskip-0.7cm
\begin{align}
	G_{+}^{BTZ}&(v_2,z_2;v_1,z_1;k) \label{eq:GfBTZ}\\
		=\,& \frac{1}{4\pi}\sqrt{\frac{z_1 z_2}{4\pi}}\!\left[\left|\Gamma\left(\frac{1}{4}\!-\!\frac{ik}{2}\right)\right|^2{}_2F_1\left(\frac{1}{4}\!-\!\frac{ik}{2},\frac{1}{4}\!+\!\frac{ik}{2},\frac{1}{2},b_1^2\right)-2b_1\left|\Gamma\left(\frac{3}{4}\!-\!\frac{ik}{2}\right)\right|^2{}_2F_1\left(\frac{3}{4}\!-\!\frac{ik}{2},\frac{3}{4}\!+\!\frac{ik}{2},\frac{3}{2},b_1^2\right)\right.\notag\\
		&\hspace{1.65cm} \left.-\left|\Gamma\left(\frac{1}{4}\!-\!\frac{ik}{2}\right)\right|^2{}_2F_1\left(\frac{1}{4}\!-\!\frac{ik}{2},\frac{1}{4}\!+\!\frac{ik}{2},\frac{1}{2},b_2^2\right)+2b_2\left|\Gamma\left(\frac{3}{4}\!-\!\frac{ik}{2}\right)\right|^2{}_2F_1\left(\frac{3}{4}\!-\!\frac{ik}{2},\frac{3}{4}\!+\!\frac{ik}{2},\frac{3}{2},b_2^2\right)\right],\notag
\end{align}
\vskip-0.2cm
\end{widetext}
where we have defined
\begin{align}
	\!\!b_1\! =& -\!z_1 z_2\!-\!(1\!-\!z_1 z_2)\cosh(v_2\!-\!v_1)\notag\\
		&-\!(z_2\!-\!z_1)\sinh(v_2\!-\!v_1)\!+\!(v_2\!-\!v_1\!+\!z_2\!-\!z_1)i\epsilon,\\
	\!\!b_2\! =&\, z_1 z_2\!-\!(1\!-\!z_1 z_2)\cosh(v_2\!-\!v_1)\notag\\
		&-\!(z_2\!-\!z_1)\sinh(v_2\!-\!v_1)\!+\!(v_2\!-\!v_1\!+\!z_2\!-\!z_1)i\epsilon.
\end{align}
The Wightman function appearing in the integral  in (\ref{eq:jformula_ini}) is between points on the AdS side of the Vaidya spacetime, where it is identical to the AdS vacuum Wightman function (\ref{eq:GfAdS}). This way, the Wightman function  between a point $(v_1,z_1)$ on the AdS side and a point $(v_3, z_3)$ on the BTZ side is given by 
\begin{align}
	G^{AB}_+(v_3,z_3;v_1,z_1;k)&\label{eq:jformula}\\
		=i \int_{v_2=0}\!\!\!\!\!\! dz_2\,&\Big[G^{AdS}_+(v_2,z_2;v_1,z_1;k)\times\notag\\
		&\,\times\!\! \overleftrightarrow{D}^{v_2}G_R^{BTZ}(v_3,z_3;v_2,z_2;k)\Big].\notag
\end{align}
We first perform the integral in (\ref{eq:jformula}) numerically using \textsc{Mathematica}'s NIntegrate. The numerical integration is somewhat subtle due to short distance singularities in the function $G_+^{AdS}$. The details on how these can be taken care of are spelled out in detail in \cite{Keranen:2014lna} and will not be repeated here.

Since finally, we are interested in the Wightman function between two points on the BTZ side, we also have to propagate the point $(v_1,z_1)$ forwards in time. For this purpose, we construct the retarded correlator between points in the AdS side and the BTZ side (where unlike previously, both of the points are located away from the $v\!=\!0$ surface) from the Wightman function using the identity
\begin{align}
G_R^{AB}&(v_3,z_3;v_1,z_1;k)\\
=&\,2i\,\theta(v_3\!+\!z_3\!-\!v_1\!-\!z_1)\Im G^{AB}_+(v_3,z_3;v_1,z_1;k).\notag
\end{align}
Thus, as we have calculated $G^{AB}_+$ numerically, we can obtain a numerical expression for the retarded
correlator as well. As a final step, we use the formula (\ref{eq:jformula}) to propagate the point $(v_1,z_1)$
forwards to the point we want using
\begin{align}
	G_+(v_3,z_3;v_4,z_4;k)&\label{eq:jformula2}\\
		=i \int_{v_1=v_1^{(0)}} dz_1\,&\Big[G^{AB}_+(v_3,z_3;v_1,z_1;k)\times\notag\\
		&\,\times\!\! \overleftrightarrow{D}^{v_1}G^{AB}_R(v_4,z_4;v_1,z_1;k)\Big],\notag
\end{align}
where $v_1^{(0)}$ is a time scale we can freely choose. The final result is expected to be independent of this choice, providing a convenient check of the numerical accuracy. Thus, the calculation finally boils
down to performing the integral in (\ref{eq:jformula2}) with the numerically known function $G_+^{AB}$.
In practice, we calculate the integral by first calculating $G^{AB}$ for a set of discrete points $z_1$
and then constructing an interpolating function from the discrete data and finally performing the numerical 
integral using NIntegrate. More details on the convergence of the numerics can be found in \cite{Keranen:2014lna}.

\section{Improved method for computing the out-of-equilibrium Wightman function}\label{sec:method_new}

Both analytical advances on the BTZ retarded propagator and improvements of the numerical methods allow us to significantly improve on the method for computing the out-of-equilibrium Wightman function presented in Appendix~\ref{sec:method_old}. The most significant progress is the idea of solving the equations of motion (\ref{eq:eoms}) for the difference $\delta G_+(z',v';z,v;k)$ rather than the Wightman function itself as outlined in Section~\ref{sec:4}.

The equations of motion (\ref{eq:eoms}) can be solved by using (\ref{eq:jformula}) for both of the
arguments of $\delta G_+$, leading to
\begin{align}
	\d G_+(v_2,z_2;&v_1,z_1;k)\label{eq:integral_2}\\
		= -\!\!\int\!\! dzdz'&\Big[G_R^{BTZ}(v_2,z_2;0,z';k)\times\notag\\
		\,&\times\! G_R^{BTZ}(v_1,z_1;0,z;k)\d \tilde{G}_+(z',z,k)\Big],\notag
\end{align}
which is equivalent to (\ref{eq:integral_1}). We introduced the notation
\begin{align}
	\d &\tilde{G}_+(z',z,k)\\
	&=\!\left(\!\frac{1}{z^2 z'^2}\!-\!\frac{2\partial_{z}}{z z'^2}\!-\!\frac{2\partial_{z'}}{z^2 z'}\!+\!\frac{4\partial_{z}\partial_{z'}}{z z'}\!\right)\d G_+(z',0;z;0,k).\notag
\end{align}
Taking the boundary limit $z_1\rightarrow 0$ and $z_2\rightarrow 0$ and using the definition of the bulk-to-boundary propagator (\ref{eq:bulk_to_boundary}), the boundary CFT correlator becomes
\begin{align}
	\d G_+^{CFT}(v_2;v_1;k)\! = \!-\!\!\int\!\! dzdz'&\Big[G_R^{bb}(v_2;0,z';k)\times\label{eq:integral_boundary}\\
		&\hspace{-0.8cm}\times\! G_R^{bb}(v_1;0,z;k)\d \tilde{G}_+(z',z,k)\Big].\notag
\end{align}
The bulk-to-boundary retarded propagator in the BTZ background can be further simplified as follows. Using the identity \cite{DLMF}
\begin{align}
	&{}_2F_1(a,b,c,x+i\epsilon)-{}_2F_1(a,b,c,x-i\epsilon)\\
	&\hspace{0.5cm}= \frac{2\pi i\, \Gamma(c)\,{}_2F_1\left(c\!-\!a,c\!-\!b,c\!-\!a\!-\!b\!+\!1,1\!-\!x\right)}{\Gamma(a)\Gamma(b)\Gamma(c\!-\!a\!-\!b\!+\!1)(x\!-\!1)^{a+b-c}}\notag
\end{align}
to extract the imaginary part of the hypergeometric functions, we obtain the BTZ retarded propagator from the Wightman function (\ref{eq:GfBTZ}),
\begin{align}
	G&{}^{BTZ}_R(v_2,z_2;v_1,z_1;k)\label{eq:GrBTZ}\\
		&= 2i\theta(v_2\!+\!z_2\!-\!v_1\!-z_1)\Im G_{+}^{BTZ}(v_2,z_2;v_1,z_1;k)\notag\\
		&= \frac{i\sqrt{z_1 z_2}}{2}\theta(v_2\!+\!z_2\!-\!v_1\!-z_1)\times\notag\\
		&\quad\; \times\left[\theta\left(z_a^*\!-\!z_1\right){\,}_2F_1\left(\frac{1}{4}\!-\!\frac{ik}{2},\frac{1}{4}\!+\!\frac{ik}{2},1,1\!-\!b_2^2\right)\right.\notag\\
		&\qquad\quad\left.-\theta\left(z_b^*\!-\!z_1\right){\,}_2F_1\left(\frac{1}{4}\!-\!\frac{ik}{2},\frac{1}{4}\!+\!\frac{ik}{2},1,1\!-\!b_1^2\right)\right],\notag
\end{align}
where
\begin{align}
	\!\!z_a^*(v_2\!-\!v_1,z_2) &= \tanh \frac{v_2-v_1}{2},\label{eq:lightcone}\\
	\!\!z_b^*(v_2\!-\!v_1,z_2) &= \frac{1\!-\!\cosh (v_2\!-\!v_1)\!-\!z_2 \sinh (v_2\!-\!v_1)}{z_2\!-\!z_2 \cosh (v_2\!-\!v_1)\!-\!\sinh (v_2\!-\!v_1)}\notag
\end{align}
mark the location of the lightcone. We can explicitly take the boundary limit, $z_2\to 0$, to obtain the bulk-to-boundary propagator. In this limit, $z_b^*\to z_a^*$ and the $\theta$-function give rise to a $\delta$-function contribution on the lightcone. We find the expression
\begin{align}
	G_R^{bb}&(v;0,z_1;k)\notag\\
		=\,&i\sqrt{2\pi}\Bigg[\frac{\left(1\!+\!4k^2\right) z_1^{\frac{3}{2}}}{8}\theta\left(z_a^*\!-\!z_1\right)\times\notag\\
		&\hspace{4mm}\times\!\! {\,}_2F_1\!\!\left(\!\frac{3}{4}\!-\!\frac{ik}{2},\frac{3}{4}\!+\!\frac{ik}{2},2,1\!-\!\left(\cosh v\!-\! z_1\sinh v\right)^2\!\right)\notag\\
		&\hspace{4mm}-\frac{\sqrt{z_a^*}}{1\!+\!\cosh v}\d\left(z_a^*\!-\!z_1\right)\Bigg].\label{eq:GRbb}
\end{align}
Combining this result with the expression (\ref{eq:integral_boundary}) for the boundary CFT Wightman function, we find a simple expression that we can integrate numerically while evaluating the $\d$-functions analytically. The resulting one- and two-dimensional integrals are performed using \textsc{Mathematica}'s NIntegrate. This way, we end up computing lower dimensional integrals than in the "old" method and avoid constructing interpolating functions. This improves the numerical accuracy, allowing us to extract the Wightman correlator until much later times. We have found that this method agrees with the old method up to the numerical accuracy of the old method. Thus, the numerical convergence results in \cite{Keranen:2014lna} apply to this method as well.

\section{Expansion of BTZ retarded propagator in terms of quasinormal modes}\label{sec:QNM}

The retarded bulk-to-boundary propagator in the BTZ black hole, given in (\ref{eq:GRbb}), has an expansion with terms falling off according to the quasinormal modes of the BTZ black hole \cite{Son:2002sd},
\begin{equation}
	\omega_*^{(n)}=\pm k-i\, 2\pi T_f(\Delta+2n).
\end{equation}
This expansion is essentially an expansion of the Gauss hypergeometric function in (\ref{eq:GRbb}) for a large argument and it can be found using the linear transformation \cite{DLMF}
\begin{align}
	&\frac{\sin(\pi(a\!-\!b))}{\pi \Gamma(c)}{}_2F_1(a,b,c,z)\\
		&\hspace{0.4cm}= \frac{(-z)^{-a}{}_2F_1(a,a\!-\!c\!+\!1,a\!-\!b\!+\!1,1/z)}{\Gamma(b)\Gamma(c\!-\!a)\Gamma(a\!-\!b\!+\!1)}-(a\!\leftrightarrow\!b)\notag
\end{align}
together with the defining series representation
\begin{equation}
	{}_2F_1(a,b,c,z)=\sum_{s=0}^\infty\frac{(a)_s(b)_s}{(c)_s s!}z^s,
\end{equation}
where $(a)_n=\Gamma(a+n)/\Gamma(a)$. The first two terms in the expansion are
\begin{widetext}
\vskip-0.7cm
\begin{align}
	iG_R^{bb}(v';v,z;k) =\,& \left(\frac{(2 z)^{\frac{3}{2}}\Gamma(-ik)e^{-ik(v'-v)}}{(1-z)^{\frac{3}{2}-ik}\Gamma\left(-\frac{1}{2}-ik\right)}+c.c\right)e^{-\frac{3}{2}(v'-v)}\notag\\
			&-\left(\frac{(2 z)^{\frac{3}{2}}(3+2(1+ik)z^2)\Gamma(-1-ik)e^{-ik(v'-v)}}{2(1-z)^{\frac{7}{2}-ik}\Gamma\left(-\frac{3}{2}-ik\right)}+c.c\right)e^{-\frac{7}{2}(v'-v)}+\mathcal{O}\left(\left(\frac{e^{-(v'-v)}}{1-z}\right)^\frac{11}{2}\right),\label{eq:Grbb_quasinormal}
\end{align}
\vskip-0.3cm
\end{widetext}
where, to simplify the notation, we have chosen units in which $T_f=1/2\pi$. The expansion parameter is  $e^{-(v'-v)}/(1-z)$ as the argument of the hypergeometric function becomes large when this parameter is small.

\section{Analytical treatment of the late time quasinormal decay}\label{sec:integral}

In this appendix, we provide details on demonstrating the quasinormal decay of the Wightman function in AdS$_3$-Vaidya with $\Delta=3/2$. First, consider the solution of the Klein-Gordon equation given smooth initial data. In the main text, we arrived at the identity (which we duplicate here for the readers convenience)
\beq
\phi^b(v,k) = \phi^b_{1}(v,k) + \phi^b_{2}(v, k),
\eeq
where
\begin{align}
	\phi_1^b(v,k) =\,& i \int_{0}^{z_a^*-\zeta} dz \,\tilde{\phi}_0(z,k) G^{bb}_R(v;0,z;k),\label{eq:generalsolApp1}\\
	\phi_2^b(v,k) =\,& i \int_{z_a^*-\zeta}^{z_a^*} dz \, \tilde{\phi}_0(z,k) G^{bb}_R(v;0,z;k),\label{eq:generalsolApp2}
\end{align}
and we have defined $\tilde{\phi}_0=-\phi_0/z^2 + 2\partial_z\phi_0/z$. So far $\zeta$ is an arbitrary positive real number. As we will see later this section, $\zeta$ should be chosen in a way that $1\gg \zeta \gg e^{-v}$. Using the quasinormal mode expansion (\ref{eq:Grbb_quasinormal}) inside the integral in (\ref{eq:generalsolApp1}), we see explicitly that this term decays at late times with a rate set by the lowest quasinormal mode. This is because the largest term in (\ref{eq:Grbb_quasinormal}) is of the order $e^{-\frac{3}{2}v}$ at late times, while the only other time dependence comes from the upper limit of the integral, $z = z_a^*-\zeta $. At late times, we can approximate $z^*_a-\zeta\approx 1-\zeta$, which leads us to conclude that
\beq
\phi^b_{1}(v, k) \propto e^{-\frac{3}{2}v}
\eeq
at late times. That is, $\phi^b_1$ decays with the lowest quasinormal mode. Higher order terms in the quasinormal mode expansion (\ref{eq:Grbb_quasinormal}) lead to a power series of the form
\beq
\phi^b_{1}(v, k)=\frac{e^{-\frac{3}{2}v}}{\sqrt{\zeta}}\sum_{n=0}^{\infty} a_n \Big(\frac{e^{-v}}{\zeta}\Big)^n.
\eeq
This implies that the quasinormal mode expansion is a good approximation when $e^{-v}/\zeta \ll 1$.

Let us consider the second term (\ref{eq:generalsolApp2}). In this integral, we cannot use the quasinormal mode expansion because it does not converge near the point $z' = z^*_a$. Instead, we will assume that $\zeta$ can be chosen to be small (while satisfying $\zeta\gg e^{-v}$ of course). This will be justified later. Also, we assume that $\phi_0(z',k)$ is a smooth function in a region near the horizon, i.e.\ we assume that all the derivatives $\partial^n_{z'}\phi_0(z',k)$ remain finite in this region. Given these two assumptions, we can first Taylor expand
\begin{align}
	z&{}^{\frac{3}{2}}\phi_0(z,k)\\
		&= (z^*_a)^{\frac{3}{2}} \phi_0(z^*_a,k) \!+\! (z\! -\! z^*_a)\partial_{z}(z^{\frac{3}{2}}\phi_0(z,k))_{z=z^*_a}+\ldots,\notag
\end{align}
and substitute into the integral in (\ref{eq:generalsolApp2}). This leads to 
\begin{align}
	\phi^b_2 & (v, k)\label{eq:phi2app}\\
		&= \Big({z^*_a}^{\frac{3}{2}}\phi_0(z^*_a,k) \!+\! \mathcal{O}(\zeta)\Big)\!\!\int_{z_a^*-\zeta}^{z_a^*}\!\frac{dz}{z^{\frac{3}{2}}} G_R^{bb}(v;0,z;k).\notag
\end{align}
The remaining integral can be performed analytically using the expression (\ref{eq:GRbb}) for the retarded bulk-to-boundary propagator,
\begin{equation}
	\!\!\!\!\int_{z_a^*-\zeta}^{z_a^*}\!\!\frac{dz}{z^{\frac{3}{2}}}G_R^{bb}(v;0,z;k)\!=\!\frac{i\cosh(\pi k)}{\sqrt{2}\pi \sinh v}F(1\!+\!\zeta \sinh v),
\end{equation}
where we have used the notation
\begin{align}
	F(x)=\,&\left|\Gamma\!\left(\!\frac{1}{4}\!-\!\frac{ik}{2}\!\right)\right|^2{\!\!\!}_2F_1\!\left(\!\frac{1}{4}\!-\!\frac{ik}{2},\frac{1}{4}\!+\!\frac{ik}{2},\frac{1}{2},x^2\!\right)\\
		&-2x\left|\Gamma\!\left(\!\frac{3}{4}\!-\!\frac{ik}{2}\!\right)\right|^2{\!\!\!}_2F_1\!\left(\!\frac{3}{4}\!-\!\frac{ik}{2},\frac{3}{4}\!+\!\frac{ik}{2},\frac{3}{2},x^2\!\right).\notag
\end{align}
For $\zeta^{-1}e^{-v}\ll 1$, we can approximate
\begin{align}
	&\frac{i\cosh(\pi k)}{\sqrt{2}\pi \sinh v}F(1+\zeta \sinh v)\label{eq:approx1}\\
	&=-i\!\left(\!\frac{2^{\frac{3}{2}}\Gamma(-ik)e^{-ikv}}{\zeta^{\frac{1}{2}+ik}\Gamma\!\left(\frac{1}{2}\!-\!ik\right)}\!+\!c.c.\!\right)e^{-\frac{3}{2}v}\!+\!\mathcal{O}\!\left(\!\left(\zeta^{-1}e^{-v}\right)^{\frac{5}{2}}\!\right).\notag
\end{align}
Substituting this result into (\ref{eq:phi2app}) leads to 
\beq
\phi^b_2(v, k)\propto e^{-\frac{3}{2}v}\Big(1 + \mathcal{O}(\zeta)\Big).\label{eq:phi2app2}
\eeq
Thus, as a second condition, the parameter $\zeta$ must be chosen so that $\zeta\ll 1$. At late times, we can satisfy both of the conditions $1\gg \zeta \gg e^{-v}$, which allows us to conclude that $\phi^b$ decays with a rate set by the lowest quasinormal mode at late times. We note that the corrections to (\ref{eq:phi2app2}) appear in two forms --- as a series expansion in powers of $e^{-v}/\zeta$ and as an expansion in derivatives of the initial data of the schematic form $\partial^n_z\phi_0 \zeta^n e^{-\frac{3}{2}v}$. The initial data has to be smooth exactly since the latter expansion in derivatives of the initial data would fail otherwise. Finally, we can employ the same approximation to the Wightman function integrals (\ref{eq:integral_boundary}), arriving at
\begin{widetext}
\vskip-0.7cm
\begin{align}
	\d G&{}_+^{CFT}(v;v;k) =\, -\int_0^{z_a^*}\! dzdz'\, G_R^{bb}(v;0,z';k) G_R^{bb}(v;0,z;k)\d \tilde{G}_+(z',z,k)\notag\\
		=\,&-\int_0^{z_a^*-\zeta}\!\!\!\!\!\!\!\!dzdz'\, G_R^{bb}(v;0,z';k) G_R^{bb}(v;0,z;k)\d \tilde{G}_+(z',z,k)-\left(\int_{z_a^*-\zeta}^{z_a^*}\frac{dz}{z^{\frac{3}{2}}}G_R^{bb}(v;0,z;k)\right)^2 \left({z_a^*}^3\d \tilde{G}_+(z_a^*,z_a^*,k)+\mathcal{O}(\zeta)\right)\notag\\
			&-2\left(\int_{z_a^*-\zeta}^{z_a^*}\frac{dz}{z^{\frac{3}{2}}}G_R^{bb}(v;0,z;k)\right)\int_0^{z_a^*-\zeta}dz'\, G_R^{bb}(v;0,z';k) \left({z_a^*}^{\frac{3}{2}}\d \tilde{G}_+(z',z_a^*,k)+\mathcal{O}(\zeta)\right).		\label{eq:separate_integrals}
\end{align}
\vskip-0.4cm
\end{widetext}
Furthermore, as seen in (\ref{eq:Grbb_quasinormal}), we can approximate the retarded bulk to boundary propagator (\ref{eq:GRbb}) as
\begin{align}
	G{}_R^{bb}(v;0,z;k) =& -\!i\!\left(\!\!\frac{(2z)^{\frac{3}{2}}\Gamma\left(-ik\right)e^{-ikv}}{(1\!-\!z)^{\frac{3}{2}+ik}\Gamma\left(\!\frac{-1-2ik}{2}\!\right)}+c.c.\!\!\right)\!e^{-\frac{3}{2}v}\notag\\
		&+\!\mathcal{O}\!\left(\left((1-z)^{-1}e^{-v}\right)^{\frac{7}{2}}\right).\label{eq:approx2}
\end{align}
Applying the approximations (\ref{eq:approx1}) and (\ref{eq:approx2}) in (\ref{eq:separate_integrals}) and employing the definitions
\begin{equation}
	iG_{\!R}^{bb,i}\!(v,\zeta,k) = \left(\!\frac{2^{\frac{3}{2}}\Gamma(-ik)e^{-ikv}}{\!\zeta^{\frac{1}{2}+ik}\Gamma\!\left(\frac{1-2ik}{2}\right)}\!+\!c.c.\!\right)\!e^{-\frac{3}{2}v}  \label{eq:Grint}\end{equation}
for the approximate integrated retarded correlator close to the lightcone and
\begin{align}
	iG_{\!R}^{bb,a}&(v,z,k)\label{eq:Grapp}\\
	&=\! \left(\!\frac{(2z)^{\frac{3}{2}}\Gamma\!\left(-ik\right)e^{-ikv}}{(1\!-\!z)^{\!\frac{3}{2}+ik}\Gamma\!\left(\!\frac{-1-2ik}{2}\right)}+c.c.\!\right)\!e^{-\frac{3}{2}v} \notag
\end{align}
for the approximation away from the lightcone, we can see that the leading term of the subtracted boundary CFT equal-time Wightman function at late time is given by
\begin{align}
	\d G&{}_+^{CFT}(v,v,k)\notag\\
		=& -\!\!\int_0^{1\!-\!\zeta}\!\!\!\!\!\!\!\! dz\,dz'\!\left[G_R^{bb,a}(v,z'\!,k)G_R^{bb,a}(v,z,k)\d \tilde{G}_+(z'\!,z,k)\right]\notag\\
		&-2G_R^{bb,i}\!(v,\zeta,k)\!\!\!\int_0^{1\!-\!\zeta}\!\!\!\!\!\!\!\! dz \left[G_R^{bb,a}\!(v,z,k) {z_a^*}^{\frac{3}{2}}\d \tilde{G}_{\!+}(z_a^*,z,k)\right]\notag\\
		&-\!G_R^{bb,i}(v,\zeta,k)^2 {z_a^*}^3\d \tilde{G}_+(z_a^*,z_a^*,k).\label{eq:integral_result}
\end{align}
The error in the approximate expression (\ref{eq:integral_result}) is again controlled by two expansions. One in powers of $\zeta^{-1}e^{-v}$, where the first subleading contribution now comes with a power of $4$. The other expansion appears with terms of the form $e^{-\frac{3}{2}v}\zeta^n \partial_z^n\left(z^{\frac{3}{2}}\d \tilde{G}_+\right)_{z=z_a^*}$, which can be truncated as $\zeta\ll1$ and the initial data is smooth. In consequence, we explicitly see that the leading late time contribution of $\d G_+^{CFT}$ falls off according to
\begin{equation}
	\delta G_+^{CFT}(v,v,k) \propto e^{-3 (2\pi T_f v)}=e^{2 v \Im \omega_*^{(0)}}
\end{equation} 
at late times, because all terms explicitly have this falloff (where we have restored $T_f$ by dimensional analysis).

\begin{figure}[t]
\begin{center}
\includegraphics[width=0.975\linewidth]{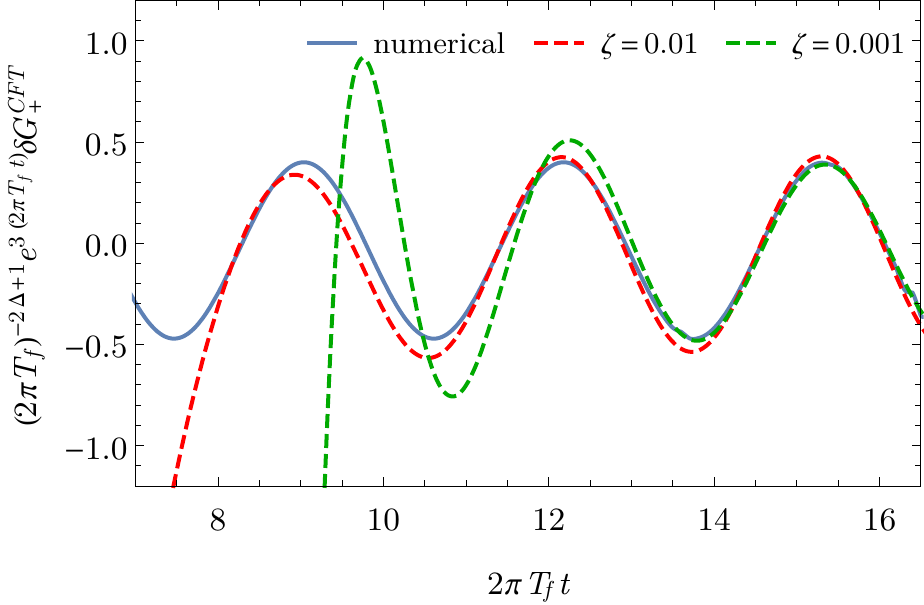}
\caption{\label{fig:QNM_approximation} 
The boundary theory Wightman function as a function of time with the thermal Wightman function subtracted and scaled to cancel the fall-off at late times, $(2\pi T_f)^{2\Delta+1} e^{3(2\pi T_f t)}\d G_+^{CFT}(t,t,k)$. The blue solid curve corresponds to the numerical result for the correlator while the red and the green curves are approximations made using the expansion of the retarded propagator in the black hole background in terms of the black hole quasinormal modes. We made two choices for the parameter $\zeta$. There is a tradeoff in the choice of $\zeta$. For smaller $\zeta$ the late time approximation is more precise, while the approximation starts to converge earlier for larger $\zeta$, as $e^{-v}\ll \zeta$ is satisfied earlier. We have also fixed $k=2\pi T_f$.}
\end{center}
\end{figure}

As a concrete test of the approximations leading to (\ref{eq:integral_result}), Fig.\ \ref{fig:QNM_approximation}
shows a comparison of (\ref{eq:integral_result}) and the full numerical integration of (\ref{eq:integral_2}),
showing good agreement at late times.

\section{Effective occupation numbers with smearing}\label{sec:a}

\begin{figure}[t]
\begin{center}
\includegraphics[width=0.975\linewidth]{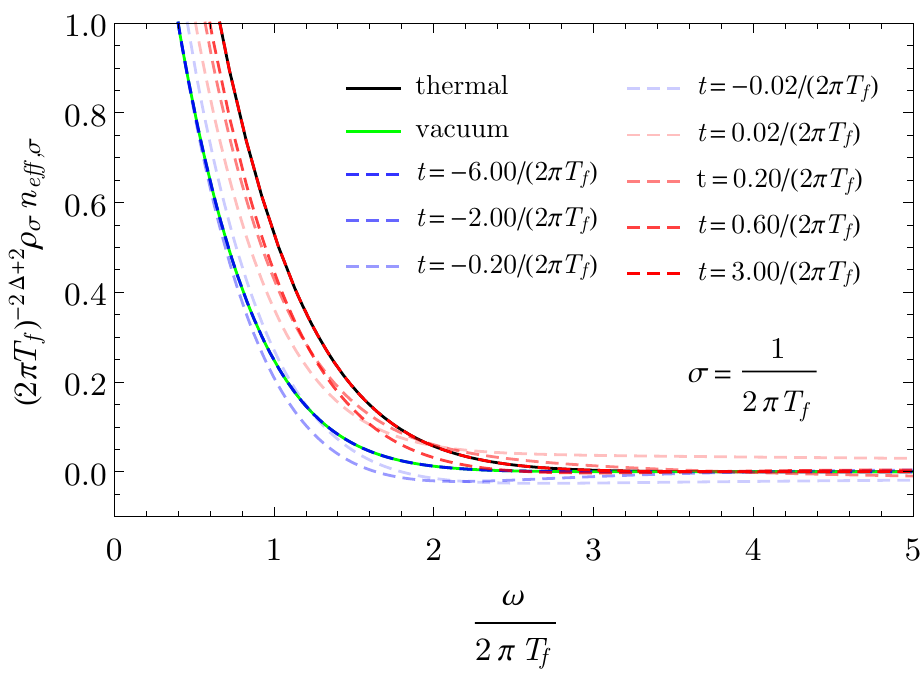}\\ \vspace{0.5cm}
\includegraphics[width=0.975\linewidth]{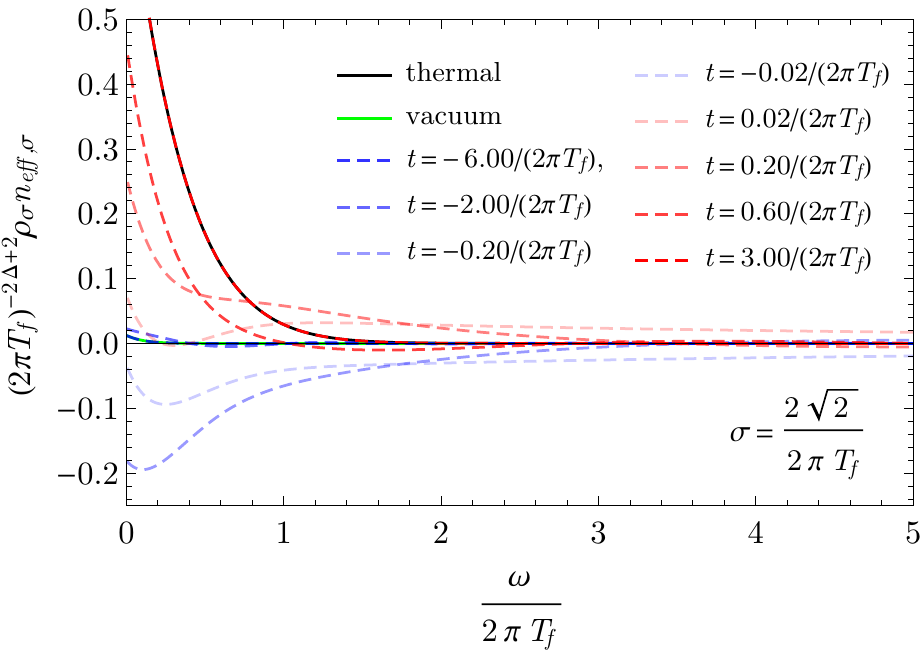}
\caption{\label{fig:n5} 
The combination $\rho_\sigma(\omega,k,t)n_{eff,\sigma}(\omega,k,t)$, which counts occupied states, for $(2\pi T_f)\sigma=1$ and $(2\pi T_f)\sigma=2\sqrt{2}$ as a function of $\omega$ for different times $t$ at momentum $k=2\pi T_f$.}
\end{center}
\end{figure}

\begin{figure}[t]
\begin{center}
\includegraphics[width=0.975\linewidth]{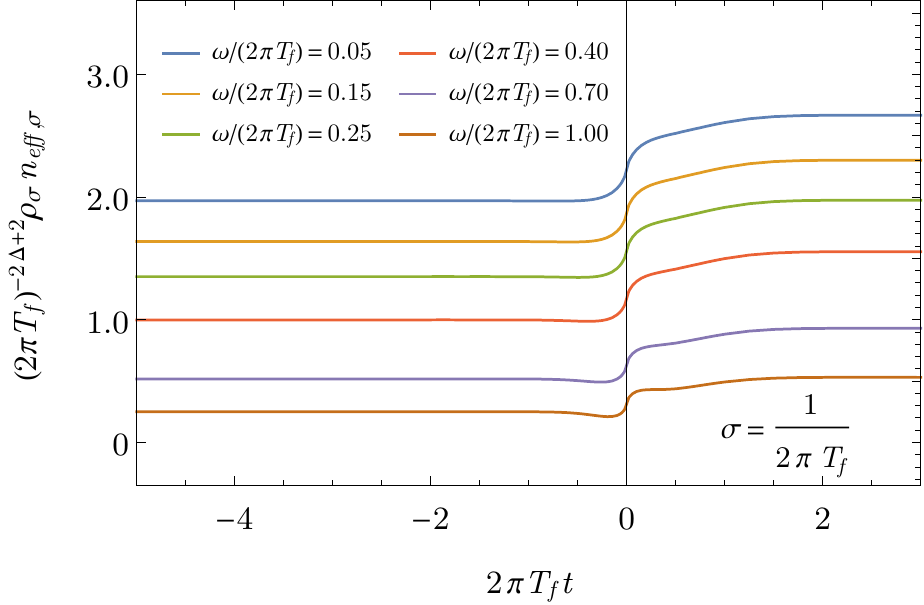}\\ \vspace{0.5cm}
\includegraphics[width=0.975\linewidth]{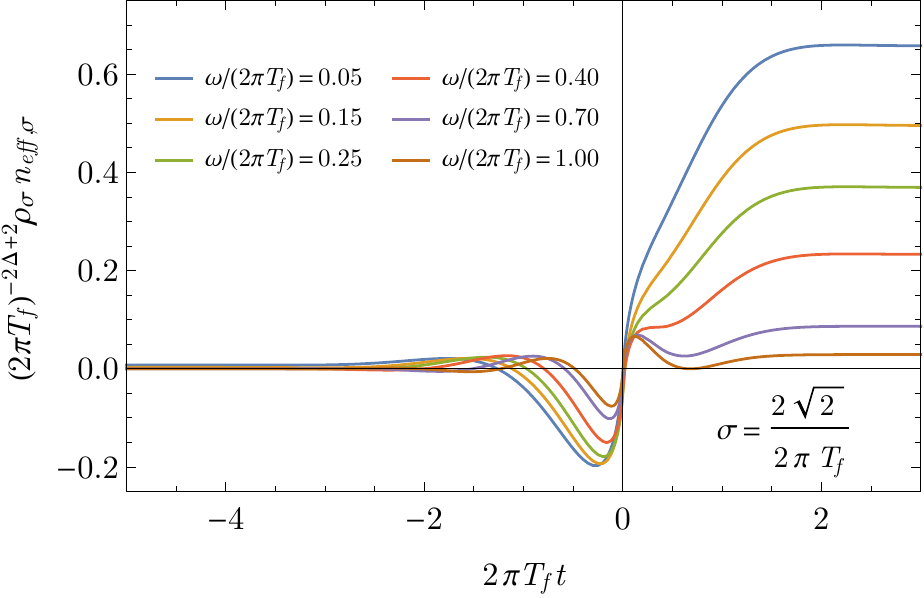}
\caption{\label{fig:n6} 
The combination $\rho_\sigma(\omega,k,t)n_{eff,\sigma}(\omega,k,t)$, which counts occupied states, for $(2\pi T_f)\sigma=1$ and $(2\pi T_f)\sigma=2\sqrt{2}$ as a function of $t$ for different energies $\omega$ at fixed momentum $k=2\pi T_f$. At early times, the effective vacuum occupation differs significantly from zero for small $\sigma$ while it is much smaller for larger $\sigma$. At late times, the effective occupation approaches the smeared thermal value. The dynamics occur in a region around the time of the quench at $t\!=\!0$ which is determined by the width of the Gaussian window in the Wigner transforms.}
\end{center}
\end{figure}

One consequence of the definition of the effective occupation number $n(\omega,k,t)$ in (\ref{eq:n}) is that it is sensitive to events far in the future and past of $t$ due to the Fourier transform with respect to relative time in the Wigner transform (\ref{eq:wigner}). This can be prevented by introducing a Gaussian smearing in the Wigner transform of the two-point function,
\begin{equation}
	G_{\sigma}(t,\omega,k)=\!\!\int\!\! d\delta t\,e^{i\omega \delta t-\frac{\delta t^2}{2\sigma^2}}G\left(\!t\!+\!\frac{\delta t}{2};t\!-\!\frac{\delta t}{2};k\!\right)\!,
\end{equation}
effectively restricting the time interval relevant to the time $t$ and thus focusing on information close to $t$. Using this smeared two-point function, we can define the effective smeared occupation number in analogy with (\ref{eq:n}) as
\beq
n_{eff,\sigma}(\omega,k,t)=\frac{G_{-,\sigma}(\omega,k,t)}{2\,\mathrm{Re}\, G_{R,\sigma}(\omega,k,t)}.\label{eq:n_smeared}
\eeq
Similarly, the corresponding spectral function is
\beq
\rho_\sigma(\omega,k,t)=2\,\textrm{Re}\,G_{R,\sigma}(\omega,k,t).
\eeq
In order to obtain the thermal smeared occupation number, we also need to introduce a Gaussian window in the thermal correlators $G_\sigma^{th}(\omega,k)$. The pure Fourier transform of the thermal correlator is known analytically and the smeared transform can be obtained from the pure one using
\begin{align}
G_{\sigma}^{th}&(\omega,k) = \!\int\!\! d\delta t\,e^{i\omega \delta t}e^{-\frac{\delta t^2}{2\omega^2}}G^{th}(t\!+\!\delta t/2,t\!-\!\delta t/2,k)\nonumber
\\
=\,& \!\int\!\!\frac{d\omega_1}{2\pi}\frac{d\omega_2}{\sqrt{2\pi}}d\delta t\,\sigma e^{i(\omega-\omega_1-\omega_2)\delta t}e^{-\frac{\sigma^2\omega_2^2}{2}}G^{th}\left(\omega_1,k\right)\nonumber
\\
=\,& \!\int\!\!\frac{d\omega_1}{\sqrt{2\pi}}\,\sigma e^{-\frac{\sigma^2(\omega-\omega_1)^2}{2}}G^{th}\left(\omega_1,k\right)
\end{align}
This integral is suitable for numerical treatment as it is sharply peaked with width $1/\sigma$ around $\omega_1\!=\!\omega$ due to the Gaussian factor.

In Fig.~\ref{fig:n5}, we show the plots of the analogies of Fig.~\ref{fig:n2} for two values of the smearing, $(2\pi T_f)\sigma=1$ and $(2\pi T_f)\sigma=2\sqrt{2}$. We can see that the smeared vacuum occupation number has a similar form as a thermal distribution and $\rho_\sigma n_{eff,\sigma}$ interpolates between this distribution and the smeared thermal distribution corresponding to the final temperature $T_f$.

The analogy of Fig.~\ref{fig:n3} for the two values of the smearing, $(2\pi T_f)\sigma=1$ and $(2\pi T_f)\sigma=2\sqrt{2}$, can be seen in Fig.~\ref{fig:n6}. For smaller $\sigma$, the dynamics of the effective occupation number becomes increasingly confined to a smaller region around $t\!=\!0$, where the quench happens, due to the Gaussian window. The Gaussian window thus makes the occupation number more local in time.

\vspace{4cm}


\bibliography{references}

\begin{thebibliography}{29}%
\makeatletter
\providecommand \@ifxundefined [1]{%
 \@ifx{#1\undefined}
}%
\providecommand \@ifnum [1]{%
 \ifnum #1\expandafter \@firstoftwo
 \else \expandafter \@secondoftwo
 \fi
}%
\providecommand \@ifx [1]{%
 \ifx #1\expandafter \@firstoftwo
 \else \expandafter \@secondoftwo
 \fi
}%
\providecommand \natexlab [1]{#1}%
\providecommand \enquote  [1]{``#1''}%
\providecommand \bibnamefont  [1]{#1}%
\providecommand \bibfnamefont [1]{#1}%
\providecommand \citenamefont [1]{#1}%
\providecommand \href@noop [0]{\@secondoftwo}%
\providecommand \href [0]{\begingroup \@sanitize@url \@href}%
\providecommand \@href[1]{\@@startlink{#1}\@@href}%
\providecommand \@@href[1]{\endgroup#1\@@endlink}%
\providecommand \@sanitize@url [0]{\catcode `\\12\catcode `\$12\catcode
  `\&12\catcode `\#12\catcode `\^12\catcode `\_12\catcode `\%12\relax}%
\providecommand \@@startlink[1]{}%
\providecommand \@@endlink[0]{}%
\providecommand \url  [0]{\begingroup\@sanitize@url \@url }%
\providecommand \@url [1]{\endgroup\@href {#1}{\urlprefix }}%
\providecommand \urlprefix  [0]{URL }%
\providecommand \Eprint [0]{\href }%
\providecommand \doibase [0]{http://dx.doi.org/}%
\providecommand \selectlanguage [0]{\@gobble}%
\providecommand \bibinfo  [0]{\@secondoftwo}%
\providecommand \bibfield  [0]{\@secondoftwo}%
\providecommand \translation [1]{[#1]}%
\providecommand \BibitemOpen [0]{}%
\providecommand \bibitemStop [0]{}%
\providecommand \bibitemNoStop [0]{.\EOS\space}%
\providecommand \EOS [0]{\spacefactor3000\relax}%
\providecommand \BibitemShut  [1]{\csname bibitem#1\endcsname}%
\let\auto@bib@innerbib\@empty
\bibitem [{\citenamefont {Bhaseen}\ \emph {et~al.}(2013)\citenamefont
  {Bhaseen}, \citenamefont {Gauntlett}, \citenamefont {Simons}, \citenamefont
  {Sonner},\ and\ \citenamefont {Wiseman}}]{Bhaseen:2012gg}%
  \BibitemOpen
  \bibfield  {author} {\bibinfo {author} {\bibfnamefont {M.}~\bibnamefont
  {Bhaseen}}, \bibinfo {author} {\bibfnamefont {J.~P.}\ \bibnamefont
  {Gauntlett}}, \bibinfo {author} {\bibfnamefont {B.}~\bibnamefont {Simons}},
  \bibinfo {author} {\bibfnamefont {J.}~\bibnamefont {Sonner}}, \ and\ \bibinfo
  {author} {\bibfnamefont {T.}~\bibnamefont {Wiseman}},\ }\href {\doibase
  10.1103/PhysRevLett.110.015301} {\bibfield  {journal} {\bibinfo  {journal}
  {Phys.Rev.Lett.}\ }\textbf {\bibinfo {volume} {110}},\ \bibinfo {pages}
  {015301} (\bibinfo {year} {2013})},\ \Eprint {http://arxiv.org/abs/1207.4194}
  {arXiv:1207.4194 [hep-th]} \BibitemShut {NoStop}%
\bibitem [{\citenamefont {Craps}\ \emph {et~al.}(2015)\citenamefont {Craps},
  \citenamefont {Lindgren},\ and\ \citenamefont {Taliotis}}]{Craps:2015upq}%
  \BibitemOpen
  \bibfield  {author} {\bibinfo {author} {\bibfnamefont {B.}~\bibnamefont
  {Craps}}, \bibinfo {author} {\bibfnamefont {E.~J.}\ \bibnamefont {Lindgren}},
  \ and\ \bibinfo {author} {\bibfnamefont {A.}~\bibnamefont {Taliotis}},\
  }\href@noop {} {\  (\bibinfo {year} {2015})},\ \Eprint
  {http://arxiv.org/abs/1511.00859} {arXiv:1511.00859 [hep-th]} \BibitemShut
  {NoStop}%
\bibitem [{\citenamefont {Abajo-Arrastia}\ \emph {et~al.}(2010)\citenamefont
  {Abajo-Arrastia}, \citenamefont {Aparicio},\ and\ \citenamefont
  {Lopez}}]{AbajoArrastia:2010yt}%
  \BibitemOpen
  \bibfield  {author} {\bibinfo {author} {\bibfnamefont {J.}~\bibnamefont
  {Abajo-Arrastia}}, \bibinfo {author} {\bibfnamefont {J.}~\bibnamefont
  {Aparicio}}, \ and\ \bibinfo {author} {\bibfnamefont {E.}~\bibnamefont
  {Lopez}},\ }\href {\doibase 10.1007/JHEP11(2010)149} {\bibfield  {journal}
  {\bibinfo  {journal} {JHEP}\ }\textbf {\bibinfo {volume} {11}},\ \bibinfo
  {pages} {149} (\bibinfo {year} {2010})},\ \Eprint
  {http://arxiv.org/abs/1006.4090} {arXiv:1006.4090 [hep-th]} \BibitemShut
  {NoStop}%
\bibitem [{\citenamefont {Albash}\ and\ \citenamefont
  {Johnson}(2011)}]{Albash:2010mv}%
  \BibitemOpen
  \bibfield  {author} {\bibinfo {author} {\bibfnamefont {T.}~\bibnamefont
  {Albash}}\ and\ \bibinfo {author} {\bibfnamefont {C.~V.}\ \bibnamefont
  {Johnson}},\ }\href {\doibase 10.1088/1367-2630/13/4/045017} {\bibfield
  {journal} {\bibinfo  {journal} {New J. Phys.}\ }\textbf {\bibinfo {volume}
  {13}},\ \bibinfo {pages} {045017} (\bibinfo {year} {2011})},\ \Eprint
  {http://arxiv.org/abs/1008.3027} {arXiv:1008.3027 [hep-th]} \BibitemShut
  {NoStop}%
\bibitem [{\citenamefont {Balasubramanian}\ \emph {et~al.}(2011)\citenamefont
  {Balasubramanian}, \citenamefont {Bernamonti}, \citenamefont {de~Boer},
  \citenamefont {Copland}, \citenamefont {Craps} \emph
  {et~al.}}]{Balasubramanian:2011ur}%
  \BibitemOpen
  \bibfield  {author} {\bibinfo {author} {\bibfnamefont {V.}~\bibnamefont
  {Balasubramanian}}, \bibinfo {author} {\bibfnamefont {A.}~\bibnamefont
  {Bernamonti}}, \bibinfo {author} {\bibfnamefont {J.}~\bibnamefont {de~Boer}},
  \bibinfo {author} {\bibfnamefont {N.}~\bibnamefont {Copland}}, \bibinfo
  {author} {\bibfnamefont {B.}~\bibnamefont {Craps}},  \emph {et~al.},\ }\href
  {\doibase 10.1103/PhysRevD.84.026010} {\bibfield  {journal} {\bibinfo
  {journal} {Phys.Rev.}\ }\textbf {\bibinfo {volume} {D84}},\ \bibinfo {pages}
  {026010} (\bibinfo {year} {2011})},\ \Eprint {http://arxiv.org/abs/1103.2683}
  {arXiv:1103.2683 [hep-th]} \BibitemShut {NoStop}%
\bibitem [{\citenamefont {Liu}\ and\ \citenamefont {Suh}(2014)}]{Liu:2013qca}%
  \BibitemOpen
  \bibfield  {author} {\bibinfo {author} {\bibfnamefont {H.}~\bibnamefont
  {Liu}}\ and\ \bibinfo {author} {\bibfnamefont {S.~J.}\ \bibnamefont {Suh}},\
  }\href {\doibase 10.1103/PhysRevD.89.066012} {\bibfield  {journal} {\bibinfo
  {journal} {Phys. Rev.}\ }\textbf {\bibinfo {volume} {D89}},\ \bibinfo {pages}
  {066012} (\bibinfo {year} {2014})},\ \Eprint {http://arxiv.org/abs/1311.1200}
  {arXiv:1311.1200 [hep-th]} \BibitemShut {NoStop}%
\bibitem [{\citenamefont {Buchel}\ \emph {et~al.}(2015)\citenamefont {Buchel},
  \citenamefont {Myers},\ and\ \citenamefont {van Niekerk}}]{Buchel:2014gta}%
  \BibitemOpen
  \bibfield  {author} {\bibinfo {author} {\bibfnamefont {A.}~\bibnamefont
  {Buchel}}, \bibinfo {author} {\bibfnamefont {R.~C.}\ \bibnamefont {Myers}}, \
  and\ \bibinfo {author} {\bibfnamefont {A.}~\bibnamefont {van Niekerk}},\
  }\href {\doibase 10.1007/JHEP07(2015)137, 10.1007/JHEP02(2015)017} {\bibfield
   {journal} {\bibinfo  {journal} {JHEP}\ }\textbf {\bibinfo {volume} {02}},\
  \bibinfo {pages} {017} (\bibinfo {year} {2015})},\ \bibinfo {note} {[Erratum:
  JHEP07,137(2015)]},\ \Eprint {http://arxiv.org/abs/1410.6201}
  {arXiv:1410.6201 [hep-th]} \BibitemShut {NoStop}%
\bibitem [{\citenamefont {Ecker}\ \emph {et~al.}(2015)\citenamefont {Ecker},
  \citenamefont {Grumiller},\ and\ \citenamefont {Stricker}}]{Ecker:2015kna}%
  \BibitemOpen
  \bibfield  {author} {\bibinfo {author} {\bibfnamefont {C.}~\bibnamefont
  {Ecker}}, \bibinfo {author} {\bibfnamefont {D.}~\bibnamefont {Grumiller}}, \
  and\ \bibinfo {author} {\bibfnamefont {S.~A.}\ \bibnamefont {Stricker}},\
  }\href {\doibase 10.1007/JHEP07(2015)146} {\bibfield  {journal} {\bibinfo
  {journal} {JHEP}\ }\textbf {\bibinfo {volume} {07}},\ \bibinfo {pages} {146}
  (\bibinfo {year} {2015})},\ \Eprint {http://arxiv.org/abs/1506.02658}
  {arXiv:1506.02658 [hep-th]} \BibitemShut {NoStop}%
\bibitem [{\citenamefont {Bhattacharyya}\ and\ \citenamefont
  {Minwalla}(2009)}]{Bhattacharyya:2009uu}%
  \BibitemOpen
  \bibfield  {author} {\bibinfo {author} {\bibfnamefont {S.}~\bibnamefont
  {Bhattacharyya}}\ and\ \bibinfo {author} {\bibfnamefont {S.}~\bibnamefont
  {Minwalla}},\ }\href {\doibase 10.1088/1126-6708/2009/09/034} {\bibfield
  {journal} {\bibinfo  {journal} {JHEP}\ }\textbf {\bibinfo {volume} {0909}},\
  \bibinfo {pages} {034} (\bibinfo {year} {2009})},\ \Eprint
  {http://arxiv.org/abs/0904.0464} {arXiv:0904.0464 [hep-th]} \BibitemShut
  {NoStop}%
\bibitem [{\citenamefont {Wu}(2012)}]{Wu:2012rib}%
  \BibitemOpen
  \bibfield  {author} {\bibinfo {author} {\bibfnamefont {B.}~\bibnamefont
  {Wu}},\ }\href {\doibase 10.1007/JHEP10(2012)133} {\bibfield  {journal}
  {\bibinfo  {journal} {JHEP}\ }\textbf {\bibinfo {volume} {1210}},\ \bibinfo
  {pages} {133} (\bibinfo {year} {2012})},\ \Eprint
  {http://arxiv.org/abs/1208.1393} {arXiv:1208.1393 [hep-th]} \BibitemShut
  {NoStop}%
\bibitem [{\citenamefont {Garfinkle}\ \emph {et~al.}(2012)\citenamefont
  {Garfinkle}, \citenamefont {Pando~Zayas},\ and\ \citenamefont
  {Reichmann}}]{Garfinkle:2011tc}%
  \BibitemOpen
  \bibfield  {author} {\bibinfo {author} {\bibfnamefont {D.}~\bibnamefont
  {Garfinkle}}, \bibinfo {author} {\bibfnamefont {L.~A.}\ \bibnamefont
  {Pando~Zayas}}, \ and\ \bibinfo {author} {\bibfnamefont {D.}~\bibnamefont
  {Reichmann}},\ }\href {\doibase 10.1007/JHEP02(2012)119} {\bibfield
  {journal} {\bibinfo  {journal} {JHEP}\ }\textbf {\bibinfo {volume} {1202}},\
  \bibinfo {pages} {119} (\bibinfo {year} {2012})},\ \Eprint
  {http://arxiv.org/abs/1110.5823} {arXiv:1110.5823 [hep-th]} \BibitemShut
  {NoStop}%
\bibitem [{\citenamefont {Horowitz}\ \emph {et~al.}(2013)\citenamefont
  {Horowitz}, \citenamefont {Iqbal},\ and\ \citenamefont
  {Santos}}]{Horowitz:2013mia}%
  \BibitemOpen
  \bibfield  {author} {\bibinfo {author} {\bibfnamefont {G.~T.}\ \bibnamefont
  {Horowitz}}, \bibinfo {author} {\bibfnamefont {N.}~\bibnamefont {Iqbal}}, \
  and\ \bibinfo {author} {\bibfnamefont {J.~E.}\ \bibnamefont {Santos}},\
  }\href {\doibase 10.1103/PhysRevD.88.126002} {\bibfield  {journal} {\bibinfo
  {journal} {Phys.Rev.}\ }\textbf {\bibinfo {volume} {D88}},\ \bibinfo {pages}
  {126002} (\bibinfo {year} {2013})},\ \Eprint {http://arxiv.org/abs/1309.5088}
  {arXiv:1309.5088 [hep-th]} \BibitemShut {NoStop}%
\bibitem [{\citenamefont {Bellac}(2011)}]{Bellac:2011kqa}%
  \BibitemOpen
  \bibfield  {author} {\bibinfo {author} {\bibfnamefont {M.~L.}\ \bibnamefont
  {Bellac}},\ }\href
  {http://www.cambridge.org/mw/academic/subjects/physics/theoretical-physics-and-mathematical-physics/thermal-field-theory?format=AR}
  {\emph {\bibinfo {title} {{Thermal Field Theory}}}}\ (\bibinfo  {publisher}
  {Cambridge University Press},\ \bibinfo {year} {2011})\BibitemShut {NoStop}%
\bibitem [{\citenamefont {Balasubramanian}\ \emph {et~al.}(2013)\citenamefont
  {Balasubramanian}, \citenamefont {Bernamonti}, \citenamefont {Craps},
  \citenamefont {Keranen}, \citenamefont {Keski-Vakkuri} \emph
  {et~al.}}]{Balasubramanian:2012tu}%
  \BibitemOpen
  \bibfield  {author} {\bibinfo {author} {\bibfnamefont {V.}~\bibnamefont
  {Balasubramanian}}, \bibinfo {author} {\bibfnamefont {A.}~\bibnamefont
  {Bernamonti}}, \bibinfo {author} {\bibfnamefont {B.}~\bibnamefont {Craps}},
  \bibinfo {author} {\bibfnamefont {V.}~\bibnamefont {Keranen}}, \bibinfo
  {author} {\bibfnamefont {E.}~\bibnamefont {Keski-Vakkuri}},  \emph {et~al.},\
  }\href {\doibase 10.1007/JHEP04(2013)069} {\bibfield  {journal} {\bibinfo
  {journal} {JHEP}\ }\textbf {\bibinfo {volume} {1304}},\ \bibinfo {pages}
  {069} (\bibinfo {year} {2013})},\ \Eprint {http://arxiv.org/abs/1212.6066}
  {arXiv:1212.6066 [hep-th]} \BibitemShut {NoStop}%
\bibitem [{\citenamefont {Berges}\ and\ \citenamefont
  {Sexty}(2011)}]{Berges:2010ez}%
  \BibitemOpen
  \bibfield  {author} {\bibinfo {author} {\bibfnamefont {J.}~\bibnamefont
  {Berges}}\ and\ \bibinfo {author} {\bibfnamefont {D.}~\bibnamefont {Sexty}},\
  }\href {\doibase 10.1103/PhysRevD.83.085004} {\bibfield  {journal} {\bibinfo
  {journal} {Phys. Rev.}\ }\textbf {\bibinfo {volume} {D83}},\ \bibinfo {pages}
  {085004} (\bibinfo {year} {2011})},\ \Eprint {http://arxiv.org/abs/1012.5944}
  {arXiv:1012.5944 [hep-ph]} \BibitemShut {NoStop}%
\bibitem [{\citenamefont {Banks}\ \emph {et~al.}(1998)\citenamefont {Banks},
  \citenamefont {Douglas}, \citenamefont {Horowitz},\ and\ \citenamefont
  {Martinec}}]{Banks:1998dd}%
  \BibitemOpen
  \bibfield  {author} {\bibinfo {author} {\bibfnamefont {T.}~\bibnamefont
  {Banks}}, \bibinfo {author} {\bibfnamefont {M.~R.}\ \bibnamefont {Douglas}},
  \bibinfo {author} {\bibfnamefont {G.~T.}\ \bibnamefont {Horowitz}}, \ and\
  \bibinfo {author} {\bibfnamefont {E.~J.}\ \bibnamefont {Martinec}},\
  }\href@noop {} {\  (\bibinfo {year} {1998})},\ \Eprint
  {http://arxiv.org/abs/hep-th/9808016} {arXiv:hep-th/9808016 [hep-th]}
  \BibitemShut {NoStop}%
\bibitem [{\citenamefont {Giddings}(1999)}]{Giddings:1999qu}%
  \BibitemOpen
  \bibfield  {author} {\bibinfo {author} {\bibfnamefont {S.~B.}\ \bibnamefont
  {Giddings}},\ }\href {\doibase 10.1103/PhysRevLett.83.2707} {\bibfield
  {journal} {\bibinfo  {journal} {Phys.Rev.Lett.}\ }\textbf {\bibinfo {volume}
  {83}},\ \bibinfo {pages} {2707} (\bibinfo {year} {1999})},\ \Eprint
  {http://arxiv.org/abs/hep-th/9903048} {arXiv:hep-th/9903048 [hep-th]}
  \BibitemShut {NoStop}%
\bibitem [{\citenamefont {Skenderis}\ and\ \citenamefont {van
  Rees}(2009)}]{Skenderis:2008dg}%
  \BibitemOpen
  \bibfield  {author} {\bibinfo {author} {\bibfnamefont {K.}~\bibnamefont
  {Skenderis}}\ and\ \bibinfo {author} {\bibfnamefont {B.~C.}\ \bibnamefont
  {van Rees}},\ }\href {\doibase 10.1088/1126-6708/2009/05/085} {\bibfield
  {journal} {\bibinfo  {journal} {JHEP}\ }\textbf {\bibinfo {volume} {0905}},\
  \bibinfo {pages} {085} (\bibinfo {year} {2009})},\ \Eprint
  {http://arxiv.org/abs/0812.2909} {arXiv:0812.2909 [hep-th]} \BibitemShut
  {NoStop}%
\bibitem [{\citenamefont {Keranen}\ and\ \citenamefont
  {Kleinert}(2015)}]{Keranen:2014lna}%
  \BibitemOpen
  \bibfield  {author} {\bibinfo {author} {\bibfnamefont {V.}~\bibnamefont
  {Keranen}}\ and\ \bibinfo {author} {\bibfnamefont {P.}~\bibnamefont
  {Kleinert}},\ }\href {\doibase 10.1007/JHEP04(2015)119} {\bibfield  {journal}
  {\bibinfo  {journal} {JHEP}\ }\textbf {\bibinfo {volume} {1504}},\ \bibinfo
  {pages} {119} (\bibinfo {year} {2015})},\ \Eprint
  {http://arxiv.org/abs/1412.2806} {arXiv:1412.2806 [hep-th]} \BibitemShut
  {NoStop}%
\bibitem [{\citenamefont {Heller}\ \emph {et~al.}(2013)\citenamefont {Heller},
  \citenamefont {Mateos}, \citenamefont {van~der Schee},\ and\ \citenamefont
  {Triana}}]{Heller:2013oxa}%
  \BibitemOpen
  \bibfield  {author} {\bibinfo {author} {\bibfnamefont {M.~P.}\ \bibnamefont
  {Heller}}, \bibinfo {author} {\bibfnamefont {D.}~\bibnamefont {Mateos}},
  \bibinfo {author} {\bibfnamefont {W.}~\bibnamefont {van~der Schee}}, \ and\
  \bibinfo {author} {\bibfnamefont {M.}~\bibnamefont {Triana}},\ }\href
  {\doibase 10.1007/JHEP09(2013)026} {\bibfield  {journal} {\bibinfo  {journal}
  {JHEP}\ }\textbf {\bibinfo {volume} {09}},\ \bibinfo {pages} {026} (\bibinfo
  {year} {2013})},\ \Eprint {http://arxiv.org/abs/1304.5172} {arXiv:1304.5172
  [hep-th]} \BibitemShut {NoStop}%
\bibitem [{\citenamefont {Louko}\ \emph {et~al.}(2000)\citenamefont {Louko},
  \citenamefont {Marolf},\ and\ \citenamefont {Ross}}]{Louko:2000tp}%
  \BibitemOpen
  \bibfield  {author} {\bibinfo {author} {\bibfnamefont {J.}~\bibnamefont
  {Louko}}, \bibinfo {author} {\bibfnamefont {D.}~\bibnamefont {Marolf}}, \
  and\ \bibinfo {author} {\bibfnamefont {S.~F.}\ \bibnamefont {Ross}},\ }\href
  {\doibase 10.1103/PhysRevD.62.044041} {\bibfield  {journal} {\bibinfo
  {journal} {Phys.Rev.}\ }\textbf {\bibinfo {volume} {D62}},\ \bibinfo {pages}
  {044041} (\bibinfo {year} {2000})},\ \Eprint
  {http://arxiv.org/abs/hep-th/0002111} {arXiv:hep-th/0002111 [hep-th]}
  \BibitemShut {NoStop}%
\bibitem [{\citenamefont {Hartman}\ and\ \citenamefont
  {Maldacena}(2013)}]{Hartman:2013qma}%
  \BibitemOpen
  \bibfield  {author} {\bibinfo {author} {\bibfnamefont {T.}~\bibnamefont
  {Hartman}}\ and\ \bibinfo {author} {\bibfnamefont {J.}~\bibnamefont
  {Maldacena}},\ }\href {\doibase 10.1007/JHEP05(2013)014} {\bibfield
  {journal} {\bibinfo  {journal} {JHEP}\ }\textbf {\bibinfo {volume} {05}},\
  \bibinfo {pages} {014} (\bibinfo {year} {2013})},\ \Eprint
  {http://arxiv.org/abs/1303.1080} {arXiv:1303.1080 [hep-th]} \BibitemShut
  {NoStop}%
\bibitem [{\citenamefont {Guica}\ and\ \citenamefont
  {Ross}(2015)}]{Guica:2014dfa}%
  \BibitemOpen
  \bibfield  {author} {\bibinfo {author} {\bibfnamefont {M.}~\bibnamefont
  {Guica}}\ and\ \bibinfo {author} {\bibfnamefont {S.~F.}\ \bibnamefont
  {Ross}},\ }\href {\doibase 10.1088/0264-9381/32/5/055014} {\bibfield
  {journal} {\bibinfo  {journal} {Class. Quant. Grav.}\ }\textbf {\bibinfo
  {volume} {32}},\ \bibinfo {pages} {055014} (\bibinfo {year} {2015})},\
  \Eprint {http://arxiv.org/abs/1412.1084} {arXiv:1412.1084 [hep-th]}
  \BibitemShut {NoStop}%
\bibitem [{\citenamefont {Callebaut}\ \emph {et~al.}(2014)\citenamefont
  {Callebaut}, \citenamefont {Craps}, \citenamefont {Galli}, \citenamefont
  {Thompson}, \citenamefont {Vanhoof}, \citenamefont {Zaanen},\ and\
  \citenamefont {Zhang}}]{Callebaut:2014tva}%
  \BibitemOpen
  \bibfield  {author} {\bibinfo {author} {\bibfnamefont {N.}~\bibnamefont
  {Callebaut}}, \bibinfo {author} {\bibfnamefont {B.}~\bibnamefont {Craps}},
  \bibinfo {author} {\bibfnamefont {F.}~\bibnamefont {Galli}}, \bibinfo
  {author} {\bibfnamefont {D.~C.}\ \bibnamefont {Thompson}}, \bibinfo {author}
  {\bibfnamefont {J.}~\bibnamefont {Vanhoof}}, \bibinfo {author} {\bibfnamefont
  {J.}~\bibnamefont {Zaanen}}, \ and\ \bibinfo {author} {\bibfnamefont {H.-b.}\
  \bibnamefont {Zhang}},\ }\href {\doibase 10.1007/JHEP10(2014)172} {\bibfield
  {journal} {\bibinfo  {journal} {JHEP}\ }\textbf {\bibinfo {volume} {10}},\
  \bibinfo {pages} {172} (\bibinfo {year} {2014})},\ \Eprint
  {http://arxiv.org/abs/1407.5975} {arXiv:1407.5975 [hep-th]} \BibitemShut
  {NoStop}%
\bibitem [{\citenamefont {David}\ and\ \citenamefont
  {Khetrapal}(2015)}]{David:2015xqa}%
  \BibitemOpen
  \bibfield  {author} {\bibinfo {author} {\bibfnamefont {J.~R.}\ \bibnamefont
  {David}}\ and\ \bibinfo {author} {\bibfnamefont {S.}~\bibnamefont
  {Khetrapal}},\ }\href {\doibase 10.1007/JHEP07(2015)041} {\bibfield
  {journal} {\bibinfo  {journal} {JHEP}\ }\textbf {\bibinfo {volume} {07}},\
  \bibinfo {pages} {041} (\bibinfo {year} {2015})},\ \Eprint
  {http://arxiv.org/abs/1504.04439} {arXiv:1504.04439 [hep-th]} \BibitemShut
  {NoStop}%
\bibitem [{\citenamefont {Chesler}\ and\ \citenamefont
  {Teaney}(2012)}]{Chesler:2012zk}%
  \BibitemOpen
  \bibfield  {author} {\bibinfo {author} {\bibfnamefont {P.~M.}\ \bibnamefont
  {Chesler}}\ and\ \bibinfo {author} {\bibfnamefont {D.}~\bibnamefont
  {Teaney}},\ }\href@noop {} {\  (\bibinfo {year} {2012})},\ \Eprint
  {http://arxiv.org/abs/1211.0343} {arXiv:1211.0343 [hep-th]} \BibitemShut
  {NoStop}%
\bibitem [{\citenamefont {Lin}(2015)}]{Lin:2015acg}%
  \BibitemOpen
  \bibfield  {author} {\bibinfo {author} {\bibfnamefont {S.}~\bibnamefont
  {Lin}},\ }\href@noop {} {\  (\bibinfo {year} {2015})},\ \Eprint
  {http://arxiv.org/abs/1511.07622} {arXiv:1511.07622 [hep-th]} \BibitemShut
  {NoStop}%
\bibitem [{DLM()}]{DLMF}%
  \BibitemOpen
  \href@noop {} {\enquote {\bibinfo {title} {{NIST Digital Library of
  Mathematical Functions}},}\ }\bibinfo {howpublished}
  {\url{http://dlmf.nist.gov/}},\ \bibinfo {note} {accessed: 07 May
  2016}\BibitemShut {NoStop}%
\bibitem [{\citenamefont {Son}\ and\ \citenamefont
  {Starinets}(2002)}]{Son:2002sd}%
  \BibitemOpen
  \bibfield  {author} {\bibinfo {author} {\bibfnamefont {D.~T.}\ \bibnamefont
  {Son}}\ and\ \bibinfo {author} {\bibfnamefont {A.~O.}\ \bibnamefont
  {Starinets}},\ }\href {\doibase 10.1088/1126-6708/2002/09/042} {\bibfield
  {journal} {\bibinfo  {journal} {JHEP}\ }\textbf {\bibinfo {volume} {0209}},\
  \bibinfo {pages} {042} (\bibinfo {year} {2002})},\ \Eprint
  {http://arxiv.org/abs/hep-th/0205051} {arXiv:hep-th/0205051 [hep-th]}
  \BibitemShut {NoStop}%
\end{thebibliography}%

\end{document}